\documentclass[%
reprint,
amsmath,
amssymb,
superscriptaddress,
aps,
]{revtex4-2}

\usepackage{pstricks}
\usepackage{graphicx}
\usepackage{dcolumn}
\usepackage{bm}
\usepackage{color}

\usepackage{braket}
\usepackage{mathrsfs}
\usepackage{siunitx}
\usepackage[]{float}
\usepackage{hyperref}

\usepackage{upgreek}
\usepackage{booktabs}
\usepackage[normalem]{ulem}

\begin{document}
	
	\preprint{APS/123-QED}
	
	\title{Benchmarking Dark Matter Search using a Parity-Check Protocol \\ with Machine-Learning Optimized Pulses}
	
	\author{Yu-Han Chang}
	\affiliation{Department of Applied Physics and InstituteQ, Aalto University, FI-00076 Aalto, Finland}
	
	\author{Ilya Moskalenko}
	\affiliation{Department of Applied Physics and InstituteQ, Aalto University, FI-00076 Aalto, Finland}
	
	\author{Marko Kuzmanović}
	\affiliation{Department of Applied Physics and InstituteQ, Aalto University, FI-00076 Aalto, Finland}
	
	\author{Ognjen Stanisavljević}
	\affiliation{Department of Applied Physics and InstituteQ, Aalto University, FI-00076 Aalto, Finland}
	
	\author{Isak Björkman}
	\affiliation{Department of Applied Physics and InstituteQ, Aalto University, FI-00076 Aalto, Finland}
	
	\author{David Díez-Ibáñez}
	\affiliation{Centro de Astropartículas y Física de Altas Energías, Universidad de Zaragoza, 50009 Zaragoza, Spain}
	
	\author{Yikun Gu}
	\affiliation{Centro de Astropartículas y Física de Altas Energías, Universidad de Zaragoza, 50009 Zaragoza, Spain}
	
	\author{Akash V. Dixit}
	\affiliation{Department of Physics, University of Chicago, Chicago, Illinois 60637, USA}
	
	\author{Igor G. Irastorza}
	\affiliation{Centro de Astropartículas y Física de Altas Energías, Universidad de Zaragoza, 50009 Zaragoza, Spain}
	
	\author{Gheorghe Sorin Paraoanu}
	\affiliation{Department of Applied Physics and InstituteQ, Aalto University, FI-00076 Aalto, Finland}
	
	
	
	\date{\today}
	
	\begin{abstract}
		We report on an improved microwave detection protocol for dark matter candidates such as the axion and the dark photon. We employ a superconducting transmon qubit dispersively coupled to a double-cavity system, enabling quantum non-demolition measurements of the photon occupation in a relatively short-lived storage cavity. To reduce the experimental cycle time and enhance sensitivity for axion and dark-photon searches, we operate this detector in a regime of increased qubit–cavity coupling, resulting in Stark shifts of 4.6 MHz. In this regime, conventional control pulses suffer from strong frequency-detuning sensitivity and photon-number–dependent errors. We address this limitation by implementing frequency-detuning–robust $\pi/2$ pulses (obtained by machine- learning optimization) that preserve high-fidelity qubit control over a bandwidth of approximately 20 MHz. We experimentally validate this protocol and demonstrate single-photon detection performance comparable to previous implementations, despite significantly reduced qubit coherence times and storage-cavity lifetimes. Using parity-based measurement sequences combined with a Hidden Markov Model (HMM) analysis, we achieve background rates on the order of $\mathcal{O}(20)$~Hz. In the absence of a magnetic field, we derive exclusion limits on the dark photon model for dark matter, reaching a sensitivity to the kinetic mixing angle of $\epsilon_{95\%} \sim 1\times10^{-14}$ at 5.051 GHz. These results establish machine-learning robust control as a key enabler for faster, more scalable microwave quantum sensors for dark-matter searches.
		
	\end{abstract}
	
	\maketitle
	
	\section{\label{sec:Intro}Introduction}
	
	The nature of dark matter (DM) remains one of the central open questions in modern physics. Although it has not yet been directly observed, its existence is strongly supported by astrophysical and cosmological evidence\cite{trimble1987existence, a2020review, gaitskell2004direct}. Among the well-motivated candidates, two prominent solutions have garnered significant attention: the \textbf{axion} and the \textbf{dark photon}. The axion, a pseudoscalar particle, was originally proposed to solve the strong-CP problem in quantum chromodynamics (QCD)~\cite{peccei1977cp, weinberg1978new, wilczek1978problem}, and its detection in laboratory haloscopes relies on the inverse Primakoff effect, in which axions convert into photons in a strong magnetic field\cite{sikivie1983experimental, graham2016vector}. The dark photon, in contrast, is a massive vector boson that interacts with the Standard Model primarily through kinetic mixing, allowing for spontaneous conversion into ordinary photons even without an external magnetic field\cite{holdom1986two, arias2012wispy, chaudhuri2015radio, Caputo2021eaa}. Both candidates can be searched for using microwave cavity haloscopes that resonantly enhance the conversion rate near the candidate frequency.
	
	Practical microwave haloscopes have historically read out the cavity field with cryogenic linear amplifiers placed at the front of the receiver chain\cite{PhysRevD.97.092001, PhysRevLett.124.101303, backes2021quantum}. However, linear amplification is fundamentally bounded by the standard quantum limit (SQL), which imposes a minimum added noise equivalent to an occupation $\bar{n}_{\text{SQL}}\!\approx\!1$ \cite{caves1982quantum}. The dark-matter--induced photon flux sits far below this floor: representative models predict an average cavity photon occupation $\bar{n}\!\lesssim\!10^{-5}$\cite{PhysRevLett.43.103, shifman1980can, Zhitnitskii1980, dine1981simple}. The only remaining handle on sensitivity requires prohibitively long integration times to achieve a sufficient signal-to-noise ratio (SNR), $t\propto SNR^2$ \cite{PhysRevD.88.035020}.
	
	The actual system noise temperature $T_{\text{sys}}$ realised in such experiments depends sensitively on where the first-stage amplifier sits in the cryogenic chain. With a HEMT at the $\!4\!$~K stage serving as the front end, microwave circuits between the cavity and the $\!4\!$~K stage add appreciable thermal noise, so that the $T_{\text{sys}}$ reaches $\sim\!7\!$~K, well above the nominal HEMT specification\cite{cervantes2024deepest}. Modern experiments such as ADMX, HAYSTAC, and CAPP therefore deploy a quantum-limited preamplifier (e.g., Josephson parametric amplifier (JPA) or travelling-wave parametric amplifier (TWPA)) directly at the mixing-chamber stage; the high gain of this cold first stage suppresses the contribution of the $\!4\!$~K HEMT, bringing $T_{\text{sys}}$ down to $0.59\pm0.31\!$ K and $0.385\pm0.025\!$ K\cite{carosi2025search, kim2024experimental}. Even in this case, however, the SQL floor of $\bar{n}_{\text{SQL}}\!\sim\!1$ still bounds how fast such a receiver can scan the parameter space.
	
	Single-photon counting with a superconducting qubit dispersively coupled to the storage cavity offers a route around the SQL bottleneck. By measuring only the cavity field amplitude---the photon number---and forgoing its conjugate phase, photon counting is no longer subject to the quantum back-action that sets the SQL\cite{PhysRevD.88.035020}. Moreover, a quantum non-demolition (QND) measurement of the photon population leaves the cavity field intact, so the same photon can be interrogated many times before it decays. The noise floor is set by the residual cavity occupation $\bar{n}_{\text{c}}$ of the cavity\cite{dixit2021searching}.
	
	A state-of-the-art realisation of this protocol has reported a residual cavity occupation $\bar{n}_{\text{c}}\!\sim\!10^{-3}$~\cite{dixit2021searching}, corresponding to an effective storage-mode temperature of a few tens of millikelvin. Reaching this floor relies on the QND character of the readout: each additional parity interrogation on the same cavity state suppresses detector-induced false counts exponentially with the number of repetitions, until the count rate is no longer limited by qubit readout errors but by the residual occupation $\bar{n}_{\text{c}}$ itself. Compared with SQL-limited receivers, the resulting noise floor lies several orders of magnitude below $\bar{n}_{\text{SQL}}\!\sim\!1$, providing the scan-rate advantage that motivates microwave single-photon counting in haloscope searches. The leverage of the protocol, however, is contingent on extracting many parity measurements within a single storage-cavity photon lifetime; the usable number of repetitions is bounded by $N\!\lesssim\!T_{1,\text{s}}\big/T_{\text{t}}$,	where $T_{1,\text{s}}$ is the photon lifetime of the storage cavity and $T_{\text{t}}$ is the duration of a single parity-measurement cycle. Maximizing $N$ therefore requires either a longer-lived cavity or a shorter measurement cycle\cite{paik2011observation, reagor2016quantum, romanenko2020three, chakram2021seamless, PRXQuantum.4.030336}.
	
	While record photon lifetimes have been demonstrated in superconducting aluminium and niobium cavities operated in zero magnetic field~\cite{romanenko2020three, PRXQuantum.4.030336, takenaka2025three, oriani2025niobium}, such conditions are not representative of axion haloscopes, which require static magnetic fields on the order of several tesla. In those environments, superconducting cavities suffer substantial degradation of their quality factors due to vortex penetration and magnetic losses~\cite{bafia2025vortex, braine2023multi, posen2023high}, so practical implementations rely on copper cavities or resonators with thin superconducting or dielectric coatings, with internal quality factors $Q_{\text{i}}\!\sim\!10^{4}\!-\!10^{5}$ in experiments such as ADMX and HAYSTAC\cite{PhysRevLett.124.101303,PhysRevD.97.092001}. Even after dedicated efforts to coat cavities with type-II superconductors~\cite{gurevich2006enhancement, posen2021advances, Ahn2022, Marconato2024, ahyoune2025rades}, the in-field quality factor remains restricted: $Q_{\text{i}}$ has been reported to decrease from $\sim\!6.6\times 10^{5}$ at 4~T to $\sim\!5.3\times 10^{5}$ at 6~T for a 3.9~GHz cavity, corresponding to photon lifetimes of $\sim\!27~\mu\mathrm{s}$ and $\sim\!22~\mu\mathrm{s}$, respectively\cite{posen2023high}. Dark-photon searches relax the strong-field requirement and can in principle reach longer $T_{1,\text{s}}$, but the practical operating regime is still set by the need to scan large frequency ranges with moderately-$Q$ tunable cavities, leaving the same incentive to reduce $T_{\text{t}}$ rather than to chase maximal $T_{1,\text{s}}$.
	
	Consequently, rather than pursuing maximally long photon lifetimes under idealized zero-field conditions, our approach targets a realistic operating regime relevant to future axion searches planned for example within the RADES collaboration. 
    In this regime, improving detector performance hinges on minimizing the duration of each parity-measurement cycle, thereby maximizing the number of QND interrogations achievable within a finite photon lifetime.
	
	Motivated by this constraint, we address the problem by optimizing quantum gate operations using \textbf{Neural Ordinary Differential Equations (Neural ODEs)}, a framework previously developed in our group\cite{kuzmanovic2025neural}. We design and experimentally validate a robust $\pi/2$ pulse that maintains high fidelity ($\mathcal{F}\!>\!99.9\%$) over a wide range of qubit--cavity detunings, enabling operation deep in the strongly Stark-shifted regime ($2\chi_{\text{qs}}\!\approx\!-4.6$~MHz) where conventional rectangular pulses fail. Our main improvement lies in the gate layer: robust Neural-ODE pulses enable the parity-measurement cycle to be shortened to $T_{\rm t}\!\sim\!1~\mu\mathrm{s}$ while maintaining $\mathcal{F}\!>\!99.9\%$ across a $\pm 20$~MHz detuning window, lifting the ceiling on $N$ under realistic, short-$T_{1,{\rm s}}$ operating conditions representative of in-field and multi-photon occupation haloscope cavities\cite{Agrawal2024}.
	
	\section{\label{sec:Methods}Methods}
	Our single-photon counter comprises three key elements: a storage cavity ($\omega_\text{s}$), a readout cavity ($\omega_\text{r}$), and a superconducting transmon qubit ($\omega_\text{q}$). The roles of the two cavities are distinct: the \textbf{readout cavity} is coupled to the external environment for efficient signal extraction, while the \textbf{storage cavity} remains isolated, interacting only with the qubit to preserve photon lifetime. Further details of the experimental setup are provided in Appendix \ref{setup}.
	
	The interaction between the transmon qubit and the cavity fields is described by the Jaynes-Cummings Hamiltonian in the dispersive regime ($|\omega_\text{r,s}-\omega_\text{q}| \gg g$). In this limit, the effective Hamiltonian takes the form:
	\begin{equation}
		H = \frac{1}{2}\omega_\text{q}\hat{\sigma}_z + (\omega_\text{r} + \chi_\text{qr}\hat{\sigma}_z)\hat{a}^{\dagger}\hat{a} + (\omega_\text{s} + \chi_\text{qs}\hat{\sigma}_z)\hat{b}^{\dagger}\hat{b}
	\end{equation}
	where $\hat{a}$ ($\hat{b}$) are the annihilation operators of the readout (storage) cavity, and $\chi_\text{qr}$ ($\chi_\text{qs}$) denote the corresponding dispersive shifts. This dispersive shift defines the photon-number-dependent qubit transition $\omega_{ge}(n)=\omega_{ge}+2\chi_{\rm qs}n$, here $2\chi_\text{qs}$ is defined as the storage–transmon Stark shift, that underlies the parity-mapping interval $T_{\rm p}$ used.
	
	\subsection{Parity measurement protocol}
	\label{sec:Protocol}
	
	\begin{figure}[t]
		\centering
		\includegraphics[width=\linewidth]{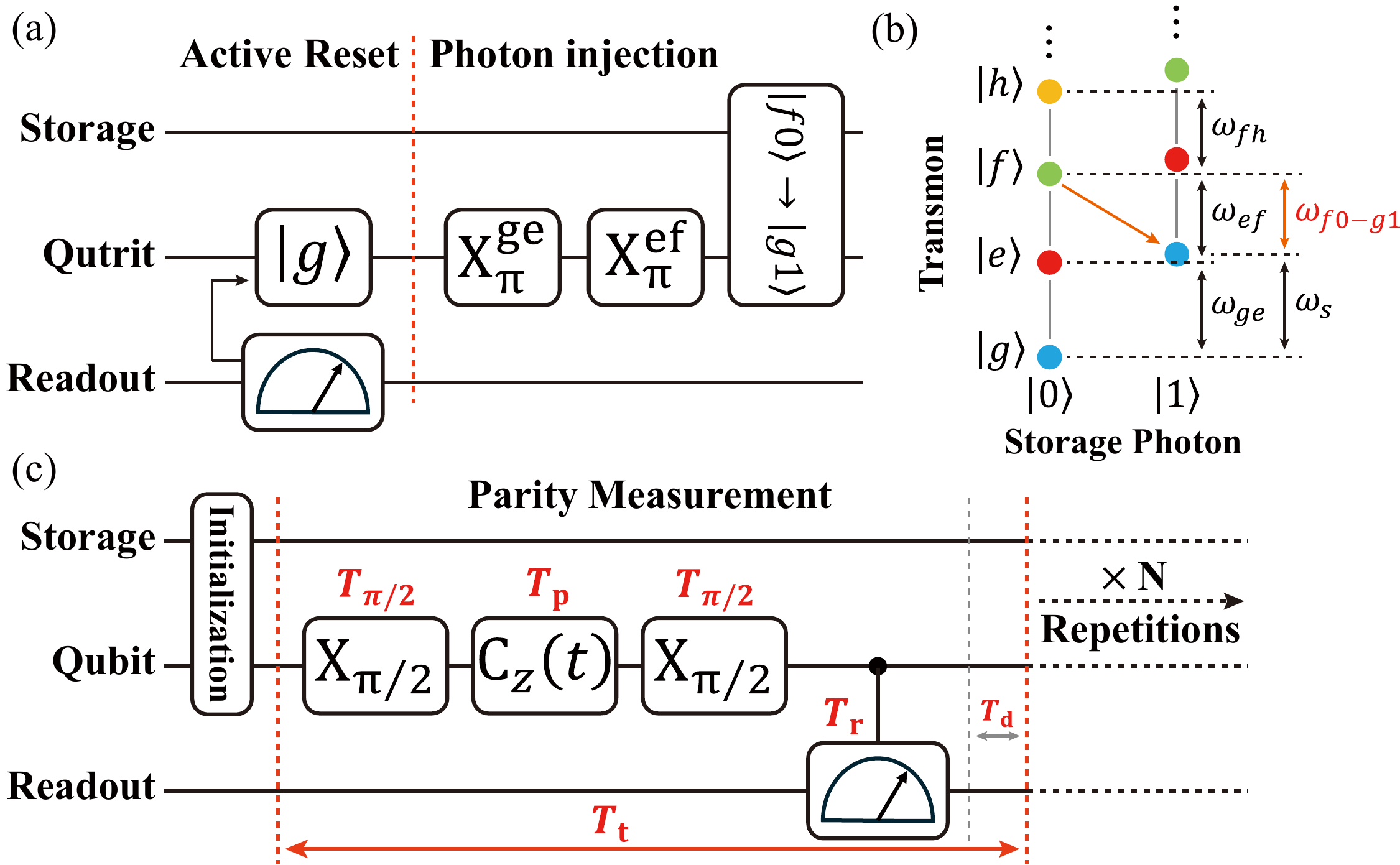}
		\caption{\textbf{Photon counter protocol.}
			\textbf{(a)} Initialization sequence applied once before each parity-measurement run: it consists of an active reset that prepares the qubit in $|g\rangle$, followed by photon injection into the storage cavity via the qubit. \textbf{(b)} Energy-level diagram of the transmon--storage system. The red arrow shows the sideband transfer $|f,0\rangle\!\rightarrow\!|g,1\rangle$	used for photon injection. \textbf{(c)} Parity-measurement cycle, repeated $N$ times after initialization. Each cycle of total duration $T_{\rm t}$ comprises two $X_{\pi/2}$ pulses of duration $T_{\pi/2}$ separated by a free-evolution interval $T_{\rm p}$, followed by a qubit readout of duration $T_{\rm r}$ and a delay $T_{\rm d}$	allowing residual readout photons to decay.}
		\label{fig:sequence}
	\end{figure}
	
	The key element of the detection scheme is the QND measurement of the photon occupation of the storage cavity mediated by the transmon qubit. This enables repeated measurements of the cavity state, thereby reducing the impact of of quantum errors in single measurements. The protocol is based on a Ramsey-type sequence that is sensitive to the storage–transmon Stark shift $2\chi_\text{qs}$, inspired by techniques originally developed in atomic physics. The pulse sequence and timing definitions are illustrated in Fig.~\ref{fig:sequence}. 
	
	Prior to each parity-measurement run, the system is initialized by a two-stage sequence shown in Fig.~\ref{fig:sequence}(a): an active reset prepares the qubit in $|g\rangle$, after which a single storage-cavity photon is prepared through the qubit by sequential $\pi$ pulses on the $\omega_{ge}$ and $\omega_{ef}$ transitions drive the qubit to $|f\rangle$, followed by an $\omega_{f0\text{-}g1}$ sideband drive that implements $|f,0\rangle\!\rightarrow\!|g,1\rangle$ the excitation to the storage cavity. The detailed	calibration of these driving fields is given in Appendix~\ref{calibration}. A single measurement cycle, with total duration $T_\text{t}$, consists of five distinct stages:
	
	\begin{enumerate}
		\item \textbf{Superposition Preparation:} The sequence begins with a $\pi/2$ pulse applied to the qubit with duration $T_{\pi /2}$, preparing the superposition state $(|g\rangle + |e\rangle)/\sqrt{2}$.
		
		\item \textbf{Parity Mapping (Free Evolution):} The qubit undergoes free evolution for a duration $T_\text{p}$, during which it accumulates a photon-number–dependent phase determined by the photon occupation $n$ of the storage cavity. The interaction time $T_\text{p}$ is calibrated such that the relative phase accumulation corresponds to a $\pi$ rotation, mapping the photon-number parity onto the qubit phase. The calibration is determined in Appendix~\ref{calibration}.4
		
		\item \textbf{State Projection:} A second $\pi/2$ pulse of duration $T_{\pi /2}$ maps the accumulated phase information into a population difference in the qubit basis ($|g\rangle$ or $|e\rangle$).
		
		\item \textbf{Readout:} The qubit state is interrogated via the readout cavity for a duration $T_\text{r}$.
		
		\item \textbf{Post-measurement delay:} A delay $T_\text{d}$ is introduced after each measurement to allow residual photons in the readout cavity to decay before the next cycle.
		
	\end{enumerate}
	
	The total cycle time is given by $T_\text{t} = 2T_{\pi /2} + T_\text{p} + T_\text{r} + T_\text{d}$. Maximizing the repetition rate therefore requires minimizing $T_\text{t}$. For the storage lifetime $T_{1,\rm s}=20.3~\mu$s measured in Appendix~\ref{calibration}.3 (Fig.~\ref{fig:T1s}), this cycle time sets the maximum number of parity cycles. We specifically target a reduction of the gate duration $T_{\pi /2}$ while preserving high gate fidelity under a large storage–transmon Stark shift $2\chi_\text{qs}$, enabled by optimal control techniques.
	
	\subsection{Robustness of $X_{\pi/2}$ gate}
	
	\begin{figure}[t]
		\centering
		\includegraphics[width=\linewidth]{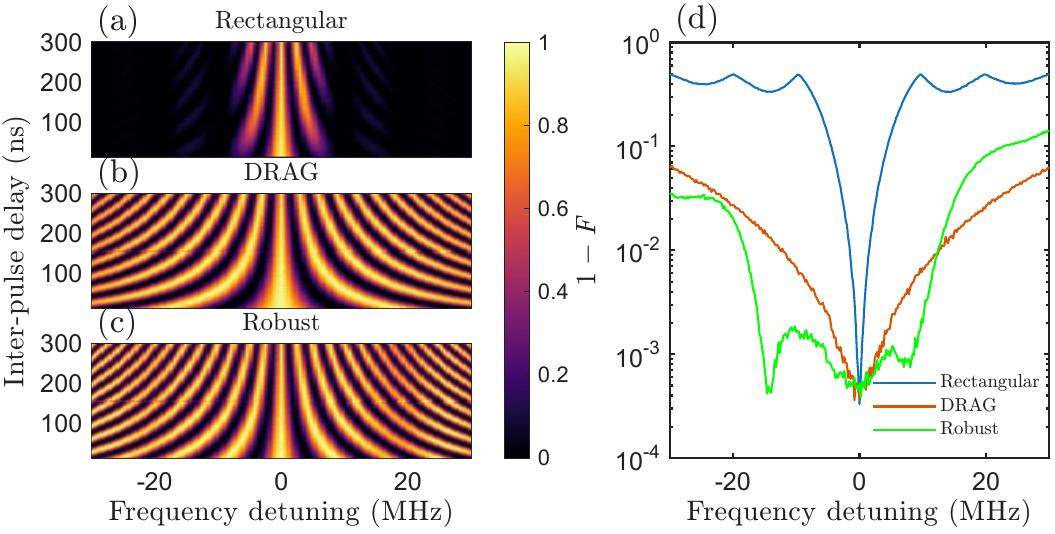}
		\caption{\textbf{Comparison of $X_{\pi/2}$ gate robustness.} All pulses share a fixed duration of 100~ns. The frequency detuning on the horizontal axis is defined as $\Delta\!=\!(\omega_{\rm drive}-\omega_{ge})/2\pi$ and is scanned over a detuning range of up to $\pm 30$~MHz.\textbf{(a)--(c)} Ramsey-fringe maps as a function of $\Delta$ and the inter-pulse delay, obtained with \textbf{(a)} a rectangular pulse, \textbf{(b)} a DRAG pulse, and \textbf{(c)} the Neural-ODE robust pulse. The color scale encodes the measured qubit excited-state population. \textbf{(d)} Gate infidelity $1\!-\!\mathcal{F}$ extracted from quantum process tomography (QPT) as a function of $\Delta$, for the rectangular (blue), DRAG (red), and robust (green) pulses. The robust pulse exhibits a nearly flat profile below $10^{-3}$ across $-15\!<\!\Delta\!<\!10$~MHz, two orders of magnitude lower than the rectangular pulse.}
		\label{fig:x90gate}
	\end{figure}
	
	In the following characterization, the qubit drive frequency is scanned over a $\pm 30$~MHz window centered on the bare qubit transition $\omega_{ge}$, so that the frequency detuning $\Delta\!\equiv\!(\omega_{\rm drive}-\omega_{ge})/2\pi$	plays the role of the photon-induced Stark shift $2n\chi_{\rm qs}$ that the gate must tolerate during a multi-photon parity readout. Since the photon counting protocol relies exclusively on the $X_{\pi/2}$ rotation of the qubit, the fidelity of this gate directly determines the operational bandwidth of the detector. Specifically, as the photon number $n$ in the storage cavity increases, the qubit transition frequency is shifted by $2n\chi_\text{qs}$. Conventional rectangular pulses fail once this photon-number–dependent shift exceeds the spectral bandwidth of the control pulse, thereby limiting the maximum detectable photon number.
	
	To quantify the improvement enabled by pulse shaping, we first perform a controlled comparison using a fixed gate duration of $T_{\pi /2} = 100$ ns for all pulse types. Fig.~\ref{fig:x90gate}(a) shows the results of Ramsey interferometry measurements.  For the rectangular pulse, rapid loss of fringe contrast is observed, with oscillations washing out beyond detunings of approximately $\pm 10$ MHz. Compared to rectangular pulse, the DRAG pulse has a significantly wider bandwidth; however, population loss can still be observed beyond $\pm 20$MHz. In contrast, the robust pulse maintains high-contrast Ramsey fringes over a substantially wider detuning range of up to $\pm 30$ MHz, demonstrating enhanced resilience to frequency shifts arising purely from waveform optimization.
	
	This robustness is further corroborated by Quantum Process Tomography (QPT), shown in Fig.~\ref{fig:x90gate}(d). The rectangular pulse (blue) is unable to implement a high-fidelity $X_{\pi/2}$ rotation across this band: its infidelity remains at the $\mathcal{O}(10^{-1})$ level throughout the scanned $\pm 30$~MHz	window, consistent with the loss of Ramsey contrast in panel (a). The DRAG pulse (red) exhibits the characteristic V-shaped profile expected from a finite-bandwidth correction, reaching $\sim\!10^{-2}$ at zero detuning and degrading by an order of magnitude near the edges of the window. By contrast, the Neural-ODE robust pulse (green) maintains a nearly flat infidelity below $10^{-3}$ across $-15\!<\!\Delta\!<\!10$~MHz, suppressing the gate error by roughly two orders of magnitude relative to the rectangular pulse over the operationally relevant Stark-shift range.
	
	While the characterization in Fig.~\ref{fig:x90gate} uses $T_{\pi/2}=100$~ns for a fair comparison against the rectangular and DRAG references, the actual photon-counting protocol runs in Sec.~\ref{sec:Benchmark} is implemented using the same dimensionless Neural-ODE waveform at a shorter $T_{\pi/2}=52$~ns. Both instances share the same on-resonance floor $1-\mathcal{F}\!\sim\!4\times 10^{-4}$, and we directly verifying the $1/T_{\pi/2}$ scaling of the operational detuning window (Appendix~\ref{sec:rescale}). This reduction in gate time provides two key advantages: it minimizes decoherence accumulated during the control operation and increases the effective control bandwidth to approximately $\pm 20$ MHz, enabling faster measurement cycles and improved multi-photon compatibility\cite{Agrawal2024}.

	\begin{table*}
		\caption{\label{tab:comparison_full} Comparison of key performance metrics for microwave single-photon detectors based on repeated QND measurements.}
		\begin{ruledtabular}
			\begin{tabular}{lcccccc}
				Detector & $T_{1,s}$ Cavity ($\mu\text{s}$) & $T_t$ Cycle Time ($\mu\text{s}$) & N (Max Meas.) & Integration time (s) & Efficiency $\eta$ & BKG Rate ($\text{sec}^{-1}$) \\ \hline
				\cite{dixit2021searching} & 546 & 10 & 30 & 8.33 & 40.3\% & 1 \\
				\cite{zhao2025flux}* & 65 & 7.3 & 25 & 10.65 & 15\% & 50 \\
				\textbf{Run 1} & $\mathbf{20.3}$ & $\mathbf{1.092}$ & $\mathbf{21}$ & $\mathbf{0.2}$ & $\mathbf{25.7\%}$ & $\mathbf{20.0}$ \\
				\textbf{Run 2} & - & - & - & - & $\mathbf{28.8\%}$ & $\mathbf{25.1}$ \\
			\end{tabular}
		\end{ruledtabular}
		\leftline{\text{* Flux-tunable storage cavity across a 22 MHz range}}
	\end{table*}
	
	\begin{figure}[b]
		\centering
		\includegraphics[width=\linewidth]{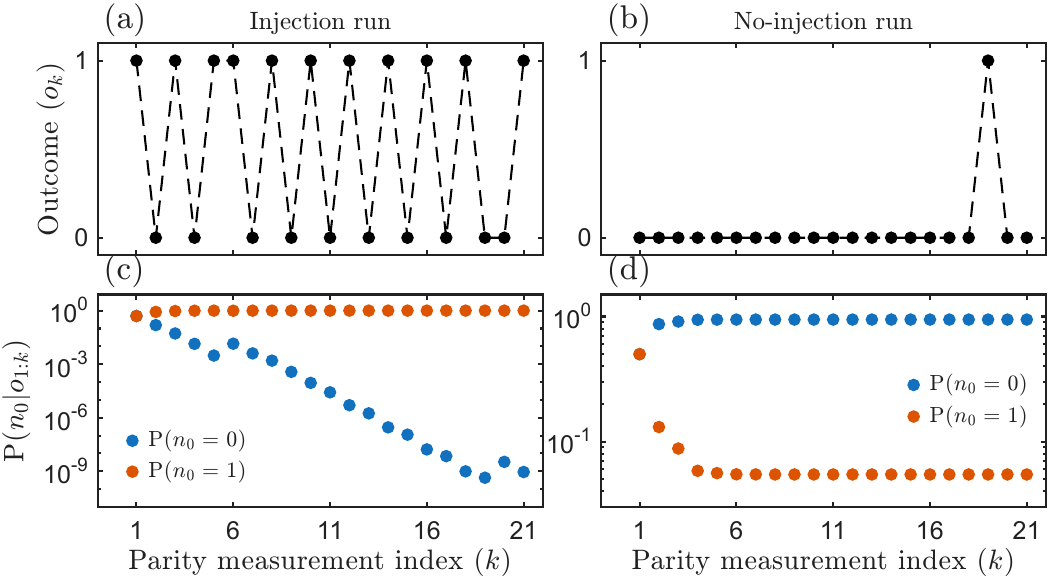}
		\caption{\textbf{The HMM analysis of parity measurements.} \textbf{(a)(b)} Two representative raw parity-measurement traces of	length $N=21$ after each cycle of the protocol. \textbf{(a)} An injection-run trace exhibiting the alternating signature of a single storage photon ($n=1$); \textbf{(b)} A no-injection-run trace showing the constant pattern characteristic of the vacuum state ($n=0$). \textbf{(c)(d)} Evolution of the HMM posterior probabilities $P(n_0\,|\,o_{1:k})$ as the number of parity measurements $k$ included in the analysis increases from $1$ to $N=21$, evaluated for the traces in panels (a) and (b), respectively.
		}
		\label{fig:ProbEvent}
	\end{figure}
	
	\subsection{Data processing}
	
	\begin{figure}[b]
		\centering
		\includegraphics[width=\linewidth]{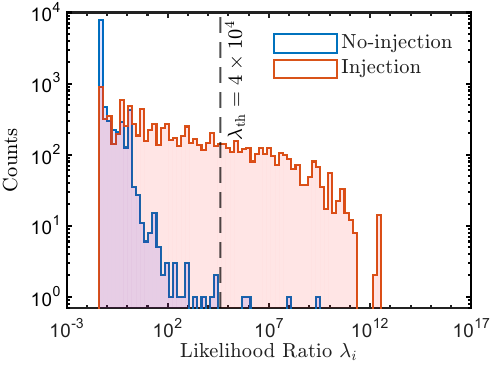}
		\caption{\textbf{The likelihood-ratio distributions.} Histograms of the per-sequence likelihood ratio $\lambda_i$ [Eq.~\eqref{eq:lambda}] evaluated by the HMM forward algorithm on $N_{\rm inj}=N_{\rm bkg}=10{,}000$ parity sequences each, for two independent datasets: the injection run (Red) and the no-injection run (blue). The	separation between the two distributions sets the single-shot discrimination performance of the detector. The vertical dashed line marks the operational threshold $\lambda_{\rm th}=4\times 10^{4}$ adopted in Sec.~\ref{sec:Benchmark}.A and discussed in Appendix~\ref{Framework}.}
		\label{fig:LikelihoodDist}
	\end{figure}
	
	The parity-measurement protocol of Sec.~\ref{sec:Protocol} yields, on each readout, a binary outcome $o\in\{g,e\}$ corresponding to the qubit state $|g\rangle$ or $|e\rangle$. A single execution of the protocol produces a parity sequence of $N=21$ such outcomes, $\mathbf{o}^{(i)}=(o^{(i)}_1,o^{(i)}_2,\ldots,o^{(i)}_N)$, indexed by the trial number $i$. All results in this section are obtained from two independent datasets of $N_{\rm inj}=N_{\rm bkg}=10{,}000$ trials each:	an \textbf{injection run}, in which loads a single photon into the storage cavity before the parity sequence begins, and	a \textbf{no-injection run}, identical in every respect except that the	injection pulse is omitted.
	
	In the ideal case, a stored photon ($n=1$) induces a $\pi$ phase shift (``phase flip'') on the qubit in each parity cycle, producing an alternating outcome sequence ($e,g,e,g,\ldots$) whose first measurement outcome is $e$; the vacuum state ($n=0$) preserves the qubit phase and yields a constant sequence ($g,g,g,\ldots$). Representative traces of both cases are shown in Fig.~\ref{fig:ProbEvent}(a)(b). In practice these signatures are distorted by qubit readout errors, finite gate fidelity, and---most importantly---stochastic decay of the storage photon over the lifetime $T_{1,\rm s}$, so that measured sequences are neither perfectly alternating nor strictly constant.
	
	To classify a noisy sequence $\mathbf{o}^{(i)}$ against the two photon-occupation hypotheses, we employ a Hidden Markov Model (HMM)~\cite{zheng2016accelerating, hann2018robust, elder2020high} whose	transition and emission probabilities are fixed by the measured storage lifetime $T_{1,\rm s}$, the readout assignment fidelity, and the per-cycle gate infidelity (Appendix~\ref{Framework}). The forward algorithm evaluates the evidence $P(\mathbf{o}^{(i)}\mid n_0)$ under each initial-state hypothesis $n_0\in\{0,1\}$; with an equal prior, Bayes' rule gives the per-sequence posterior $P(n_0\mid\mathbf{o}^{(i)})$, from which we define the per-sequence likelihood ratio:
	\begin{equation}
		\lambda_i \;\equiv\; \frac{P(n_0=1\mid\mathbf{o}^{(i)})}{P(n_0=0\mid\mathbf{o}^{(i)})}\;=\;\frac{P(\mathbf{o}^{(i)}\mid n_0=1)}{P(\mathbf{o}^{(i)}\mid n_0=0)}.
		\label{eq:lambda}
	\end{equation}
	Trial $i$ is classified as a photon detection event if $\lambda_i\!\geq\!\lambda_{\rm th}$ and as a background count otherwise. Equation~\eqref{eq:lambda} is evaluated on the full $N$-outcome sequence; the partial forward posteriors $P(n_0\mid o^{(i)}_{1:k})$ for $k \leq N$ are intermediate quantities of the HMM, shown in Fig.~\ref{fig:ProbEvent}(c)(d).
	
	Figure~\ref{fig:LikelihoodDist} shows the histogram of $\lambda_i$ for the two datasets introduced above. The injection run (red) is strongly weighted toward large $\lambda_i$, the no-injection run (blue) toward small $\lambda_i$; the separation between the two distributions sets the single-shot discrimination capability of the photon counter. Thresholding $\lambda_i$ at $\lambda_{\rm th}$ converts these per-trial decisions into the time-averaged detector metrics---the detection efficiency $\eta(\lambda_{\rm th})$ and the background rate $R_{\rm bkg}(\lambda_{\rm th})$---reported in	Sec.~\ref{sec:Benchmark}.A; the explicit mapping and the threshold-selection criterion are derived in Appendix~\ref{Framework}.
	
	\section{Benchmark Results}
	\label{sec:Benchmark}
	
	\begin{figure*}
		\centering
		\includegraphics[width=\linewidth]{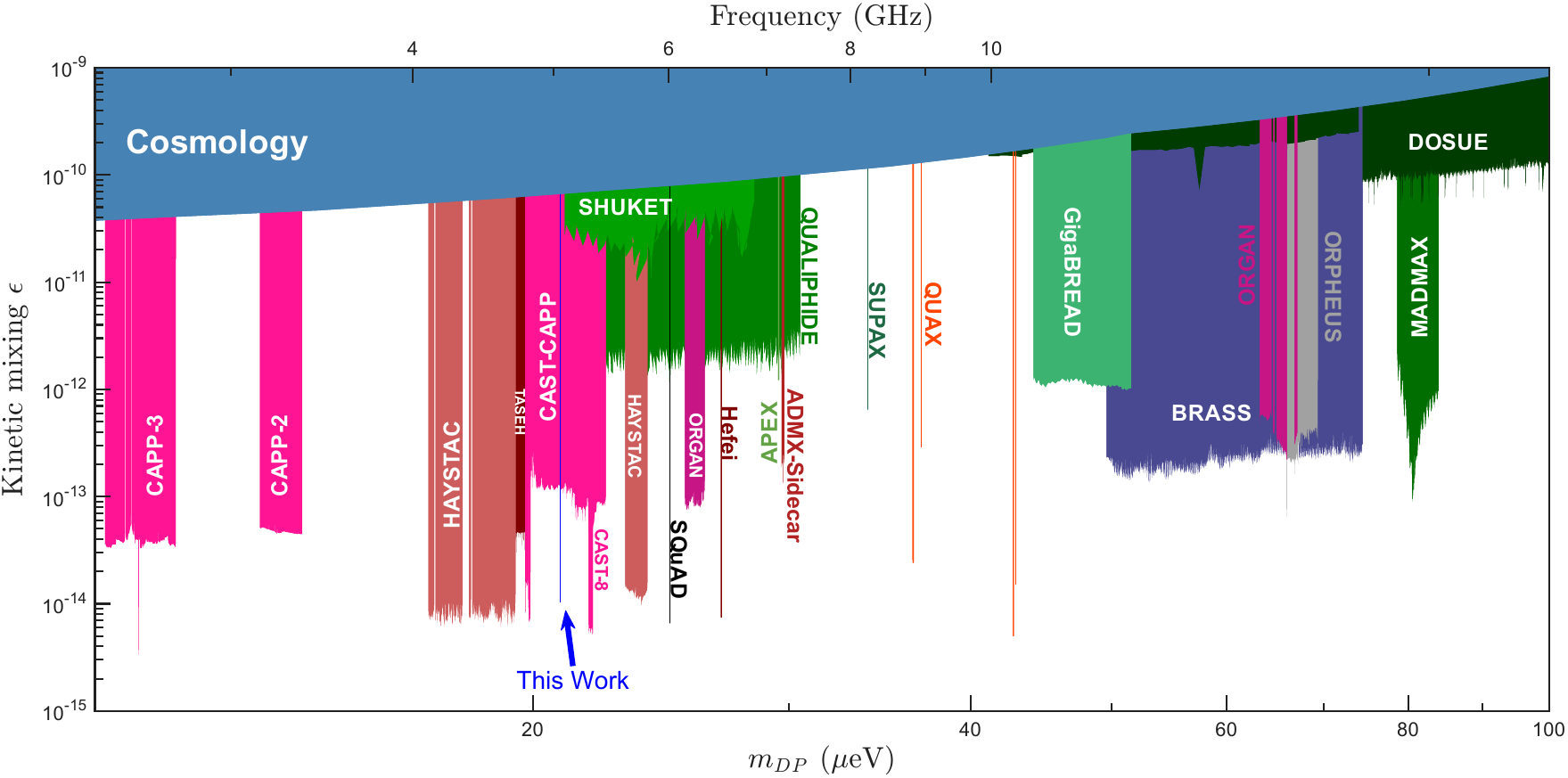}
		\caption{\textbf{Exclusion limit for dark photons.} 95\% C.L. limit derived from the quantum single-photon counter (blue), using Runs 1 and 2. Previous experimental results are shown for comparison. Plot adapted with current constraints (last update: Jan 2026). The collected limits, references and plots are available in the git-hub repository: \url{https://cajohare.github.io/AxionLimits/} (accessed on 22 June 2026).\cite{Caputo2021eaa,batkovic2021axion,AxionLimits}}
		\label{fig:DarkPhotonSensitivity}
	\end{figure*}
	
	\subsection{Detector Characterization}
	
	To evaluate the detector performance, we conducted two independent measurement runs. Each run consists of $N_{\rm inj}=10{,}000$ \textbf{injection-run} sequences and $N_{\rm bkg}=10{,}000$ \textbf{no-injection-run} sequences. The injection run provides the detection efficiency $\eta(\lambda_{\rm th})$, and the no-injection run provides the residual photon number $\bar n_c (\lambda_{\rm th})$ and the equivalent background rate $R_{\rm bkg}(\lambda_{\rm th})$. The present cavity has no additional port directly addressing the storage mode (Appendix~\ref{setup}); $\bar n_c$ is therefore obtained self-consistently from the no-injection run rather than from a displacement-based reference, and its value is cross-checked against the Bose--Einstein occupation expected from the transmon effective temperature (Appendix~\ref{Framework}).
	
	The operational threshold $\lambda_{\rm th}$ is selected from the scan of Fig.~\ref{fig:BackgroundRates}: it is placed at the onset of the plateau in $R_{\rm bkg}(\lambda_{\rm th})$, where further increases of $\lambda_{\rm th}$ no longer suppress the background rate. Beyond this point the residual triggers populate the high-$\lambda_i$ tail of the no-injection distribution [Fig.~\ref{fig:LikelihoodDist}, blue] and	cannot be removed by post-selection; they instead set the floor at which the detector is genuinely background-limited rather than signal-tail-limited. At $\lambda_{\rm th}=4\times 10^{4}$ this trade-off yields a detection efficiency $\eta\!\approx\!25.7\%$ (Run~1) and $28.8\%$ (Run~2), with bare background rates $R_{\rm bkg}^{bare}\!\approx\!20.0$ and $25.1$ $\text{sec}^{-1}$ corresponding to per-run residual cavity occupations	$\bar n_{c}\!\approx\!1.6\times 10^{-3}$ and $1.8\times 10^{-3}$ [Eq.~\eqref{eq:nbar_c}], respectively, integrated over the $0.2$~s live time per run. The two run agree on $\bar n_{c}$ to within statistical uncertainty at the per-run sample size.
	
	The averaged residual occupation $\bar n_{c}\!\approx\!1.7\times 10^{-3}$ corresponds, via the Bose--Einstein relation [Eq.~\eqref{eq:Tceff}], to an effective	storage-mode temperature $T_{\rm eff,s}\!\approx\!38$~mK, which lies between the mixing-chamber stage temperature $T_{\rm MC}\!\approx\!12$~mK and the transmon effective temperature $T_{\rm eff,q}\!\approx\!72$~mK. The decomposition of $\bar n_{c}$ into a transmon-mediated contribution $\bar n_{c}^{q}\!\sim\!6.7\times 10^{-4}$	[Eq.~\eqref{eq:ncq}, $\sim\!39\%$ of the measured $\bar n_{c}$] and an extraneous-bath contribution accounting for the remaining $\sim\!61\%$ is derived in Appendix~\ref{sec:nbar_sources}.
	
	Table~\ref{tab:comparison_full} summarizes the performance of our prototype compared to state-of-the-art QND-based microwave photon	detectors. The principal advantage of our system lies in the combination of a comparable detection efficiency ($\sim\!27\%$) and high gate fidelity under large frequency detuning, which enables operation with a higher photon number. Furthermore, the $T_{\pi/2}=52$~ns robust gate allows a comparable number of measurement cycles ($N=21$) within the relatively short cavity lifetime ($T_{1,\rm s}=20.3~\mu$s), validating our methodology of minimizing gate duration to maximize $N$.
	
	\subsection{Exclusion Limits on Massive Dark Photons}
	
	Since the measurements were performed without an applied magnetic field, the dataset constrains massive dark photons (DP), whose kinetic mixing with the cavity mode is independent of any external
	field\cite{Caputo2021eaa, chaudhuri2015radio}. Resonant conversion occurs when the dark-photon  frequency matches the storage mode, $m_{\mathrm{DP}}c^{2}=\hbar\omega_{s}$, which for our cavity corresponds to $m_{\mathrm{DP}}\simeq 20.89~\mu\mathrm{eV}$ ($\omega_{s}/2\pi=5.051$~GHz).
	
	Using the statistical framework of Appendix~\ref{Framework} [Eqs.~\eqref{eq:binom-pmf}--\eqref{eq:cdf}], we set the 95\% confidence-level (C.L.) upper limit on the kinetic mixing angle $\epsilon$ by combining the two independent data sets through Eq.~\eqref{eq:Pcomb}. The resulting exclusion curve is shown in Fig.~\ref{fig:DarkPhotonSensitivity}; the per-run inputs ($N_{\mathrm{obs}},\eta,\bar n_{\mathrm{sig}}^{95\%}$) and outputs ($\epsilon^{95\%}$) are listed in Table~\ref{tab:summary}. The	combined limit reaches $\epsilon^{95\%}\simeq 1.04\times 10^{-14}$ at $m_{\mathrm{DP}}=20.89~\mu\mathrm{eV}$, extending the search into parameter space previously unexplored by comparable cavity-based microwave setups operating in this frequency band.
	
	\section{Discussion}
	\label{sec:Discussion}
	The detector performance demonstrated in this work should be interpreted in the context of realistic operating conditions for axion haloscope experiments. While significantly longer photon
	lifetimes have been reported in superconducting cavities operated in zero magnetic field, such conditions are not compatible with axion detection, where multi-tesla static magnetic fields are required. Under these conditions, cavity quality factors are known to degrade substantially~\cite{bafia2025vortex, braine2023multi, posen2023high}, resulting in photon lifetimes that are typically limited to the tens-of-microseconds regime.
	
	From this perspective, the storage lifetime realized in our	experiment ($T_{1,\rm s}=20.3~\mu$s) is not an incidental limitation but is representative of the performance expected in practical haloscope implementations, including copper cavities and coated resonators operated in strong magnetic fields. In this regime, further increases in sensitivity cannot rely solely on extending cavity lifetimes, but instead require maximizing the information extracted per unit time.

	Our results demonstrate that this can be achieved by reducing the duration of the parity-measurement cycle rather than by pursuing marginal gains in $T_{1,\rm s}$. By operating in a strong dispersive coupling regime with a Stark shift of $2\chi_{\rm qs}/2\pi=-4.6$~MHz and employing Neural-ODE-optimized $X_{\pi/2}$ pulses of duration $T_{\pi/2}=52$~ns that retain $1-\mathcal{F}\!<\!10^{-3}$ across a $\pm 20$~MHz detuning window (Sec.~\ref{sec:Methods}.B), we complete each parity-measurement cycle in $T_{\rm t}=1.092~\mu$s and fit $N=21$ QND interrogations within a single storage lifetime. This shifts the operating bottleneck away from cavity engineering and onto the	control layer, where the present platform has substantial remaining headroom.
	
	The principal advantage of QND photon counting over linear-amplifier readout is the elimination of the standard quantum limit (SQL) as a noise floor. Phase-preserving linear amplification carries an unavoidable added noise of $\bar n_{\rm SQL}\!\sim\!1$ photon per cavity mode~\cite{caves1982quantum}, against which any dark-matter signal at $\bar n\!\lesssim\!10^{-5}$ must be discriminated by long integration. The residual cavity occupation of our detector at the operating threshold, $\bar n_{c}\!\approx\!1.7\times 10^{-3}$ [Eq.~\eqref{eq:nbar_c}, Sec.~\ref{sec:Benchmark}.A], sits roughly	three orders of magnitude below this floor:
	\begin{equation}
		\frac{\bar n_{\rm SQL}}{\bar n_{c}}\;\approx\; \frac{1}{1.7\times 10^{-3}}\;\sim\; 6\times 10^{2}.
		\label{eq:SQL_ratio}
	\end{equation}
	
	The decomposition of $\bar n_{c}$ carried out in Appendix~\ref{sec:nbar_sources} clarifies the engineering response required to push the detector deeper into the sub-SQL regime. The transmon-mediated dispersive-dressing contribution $\bar n_{c}^{q}\!\simeq\!6.7\times 10^{-4}$ [Eq.~\eqref{eq:ncq}] accounts for $\sim\!39\%$ of the measured $\bar n_{c}$ and is set by the geometric ratio $(g_{\rm qs}/\Delta)^{2}\!\simeq\!1.4\times 10^{-2}$ together with the residual transmon population $\bar n_{q}\!\approx\!4.8\times 10^{-2}$. Reducing this	component therefore acts on the qubit side: improved infrared filtering at the	qubit chip to suppress quasiparticle generation, and---in field-tolerant variants---gap engineering of the superconducting electrode that hosts the Josephson junction. Increasing $|\Delta|$ provides a complementary handle, but only at the cost of reducing	$\chi_{\rm qs}$ and thereby the parity-measurement contrast, so the present operating point already trades a modest dressing penalty for a strong $2\chi_{\rm qs}/2\pi=-4.6$~MHz Stark shift.
	
	The remaining $\sim\!61\%$ of $\bar n_{c}$ is attributed to extraneous thermal channels not mediated by the qubit (Appendix~\ref{sec:nbar_sources}). The dominant suppression levers for this component are environmental rather than qubit-level: additional infrared filters and cryogenic attenuators on the input lines, improved magnetic and electromagnetic shielding of the sample stage. The fact that the two contributions are of comparable magnitude implies that neither lever, applied in isolation, can reduce $\bar n_{c}$ by an order of magnitude; a coordinated	suppression on both fronts is required, and is the explicit subject of follow-up work.
	
	Importantly, the gate-layer strategy underlying this performance remains effective in the presence of large photon-number--dependent frequency shifts, where conventional rectangular or DRAG-based pulses suffer severe fidelity degradation (Fig.~\ref{fig:x90gate}). The use of Neural-ODE-optimized pulses	therefore provides a practical route to high-bandwidth, high-fidelity operation that is compatible with the constraints imposed by strong magnetic fields. Taken together, these results indicate that	optimizing quantum control and measurement speed---rather than	maximizing cavity lifetime under idealized conditions---is a viable	and scalable pathway toward quantum-enhanced axion and dark-photon searches. The approach demonstrated here is directly applicable to	future haloscope experiments operating in realistic magnetic-field environments and provides a complementary direction to ongoing efforts focused on cavity materials and surface treatments.
	
	\section{Summary}
	In this work, we have demonstrated a prototype single-photon counter for dark-matter searches in the microwave domain, based on a superconducting transmon qubit coupled to a 3D aluminum cavity system with storage lifetime $T_{1,\rm s}=20.3~\mu$s. By integrating Neural-ODE-optimized control pulses of duration $T_{\pi/2}=52$~ns, we implemented a robust QND parity-measurement protocol that retains gate infidelity $1-\mathcal{F}\!<\!10^{-3}$ across a $\pm 20$~MHz detuning window. This broad operational bandwidth provides two benefits: it accommodates the large Stark shifts associated with multi-photon storage states ($2\chi_{\rm qs}/2\pi=4.6$~MHz), thereby extending the detector's dynamic range beyond the single-photon	regime, and it shortens the parity-measurement cycle to	$T_{\rm t}=1.092~\mu$s, fitting $N=21$ QND interrogations within a single storage lifetime at a detection efficiency of $\eta\!\approx\!27\%$.
	
	Operating in the absence of an applied magnetic field, the prototype exhibits an averaged residual cavity occupation $\bar n_{c}\!\approx\!1.7\times 10^{-3}$ per cycle, corresponding to the bare background rate $R_{\rm bkg}^{bare}\!\approx\!22~\text{sec}^{-1}$ and an effective storage-mode temperature $T_{\rm eff,s}\!\approx\!38$~mK---roughly three orders of magnitude below the standard-quantum-limit noise floor of $\bar n_{\rm SQL}\!\sim\!1$ photon per mode. Combining the two independent data sets, we set a 95\,\% C.L.~upper limit on the kinetic-mixing parameter of $\epsilon^{95\%}=1.04\times 10^{-14}$ at the storage-cavity frequency $m_{\rm DP}c^{2}=\hbar\omega_{\rm s}$, corresponding to a dark-photon mass of $20.89~\mu$eV ($\omega_{\rm s}/2\pi=5.051$~GHz). Although obtained at zero field, the same protocol applies to axion haloscope conditions; in particular, the storage Compton frequency falls within the axion mass window currently targeted by tunable haloscope programmes, establishing this platform as a building block for QCD-axion searches at masses near $20~\mu$eV.
 	
	A microscopic decomposition of $\bar n_{c}$ attributes $\sim\!39\%$ of the residual occupation to a transmon-mediated dispersive-dressing contribution and the remaining $\sim\!61\%$ to extraneous thermal channels, placing the detector in an intermediate regime in which	the two routes contribute comparably. The remaining sensitivity	headroom is therefore addressable simultaneously at the qubit and the environmental layer of the apparatus rather than through colder operating temperatures, and future work will pursue both: improved infrared filtering at the qubit chip to suppress $\bar n_{c}^{q}$, together with additional cryogenic	attenuation and electromagnetic shielding to suppress the extraneous	contribution. Shortening the qubit readout time	$T_{\rm r}=800$~ns---currently the dominant component of $T_{\rm t}$---will further increase the number of QND interrogations achievable per storage lifetime. Together with the scalability and field tolerance of the optimized control scheme, this platform provides a promising pathway toward scanning haloscopes with sensitivity approaching the QCD-axion model predictions.
	
	\begin{acknowledgments}
		Our research has been performed as part of the  RADES collaboration, and in particular we are grateful to Laura Segui, Saiyd Ahyoune, and Jordi Miralda-Escudé for stimulating interactions and comments at different stages of this project.
		
		This work is supported by ERC Synergy Grant DarkQuantum (Grant agreement ID: 101118911, doi: 10.3030/101118911), under the programme Horizon.1.1 - European Research Council (ERC).This publication is also part of QuantERA project QRADES funded by  Agencia Estatal de Investigación project PID2019-108122GB-C31 financed by MCIU/AEI/10.13039/501100011033), by Research Council of Finland project no. 36133,  and co-funded by the European Union.
		
		The Zaragoza team acknowledges additional support from the Spanish Agencia Estatal de Investigación under grant PID2019-108122GB-C31 funded by MCIN/AEI/10.13039/501100011033. as well as support from the “European Union NextGenerationEU/PRTR” (Planes complementarios, Programa de Astrof\'isica y F\'isica de Altas Energías).  The Aalto team thanks the Jane and Aatos Erkko Foundation for funding under the doctoral program managed by InstituteQ.
		
		We would like to acknowledge the use of Servicio General de Apoyo a la Investigación-SAI, Universidad de Zaragoza. We acknowledge the provision of facilities by Aalto University at OtaNano - Micronova Nanofabrication Centre and Low Temperature Laboratory. We thank Ciaran O'Hare for his contribution to integrating experimental data from research communities.
		
	\end{acknowledgments}
	
	\bibliography{references}
	
	\clearpage
	\appendix
	
	\renewcommand{\thefigure}{S\arabic{figure}}
	\setcounter{figure}{0}
	
	\section{Experimental setup}\label{setup}
	\subsection{The cavity}
	
\begin{figure}[h]
		\centering
		\includegraphics[width=0.85\linewidth]{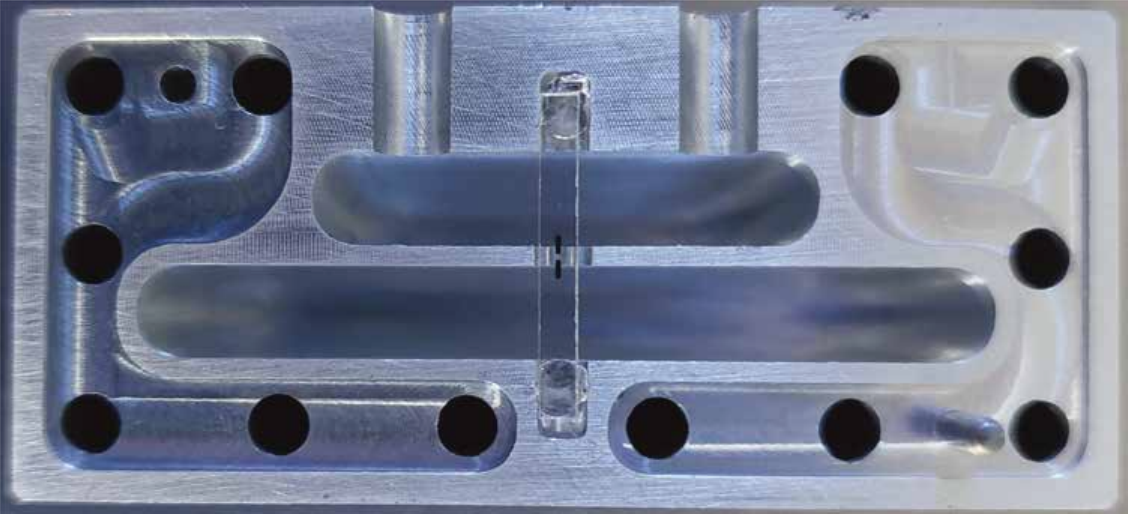}
		\caption{Photograph of one half of the aluminum 3D cavity assembly. The superconducting transmon qubit (black structure) is fabricated on a sapphire substrate and mounted in the central trench, bridging the readout cavity (top) and the storage cavity (bottom).}
		\label{fig:Al_cav}
	\end{figure}

	The 3D cavity was machined from a solid block of aluminum (alloy AW-6082; nominal purity 94–96\%) and consists of two physically distinct chambers: a readout cavity and a storage cavity. The dimensions of both cavities are summarized in Table \ref{tab:CavDims}. A narrow trench milled between the two chambers accommodates a superconducting transmon qubit that bridges them; no direct electromagnetic coupling between the cavities was implemented. Fig.~\ref{fig:Al_cav} shows one half of the assembly with the transmon installed. The sapphire substrate hosting the qubit was mechanically secured within the trench using a small amount of indium. The storage cavity is designed to maximize photon lifetime, with a fundamental mode at \(\omega_\text{s} = 2\pi \times 5.051~\mathrm{GHz}\) and a measured quality factor of \(Q_\text{s} = 6.35 \times 10^{5}\), corresponding to a photon decay time of \(T_\text{1,s} \approx 20~\mu\text{s}\). The readout cavity operates at \(\omega_\text{r} = 2\pi \times 6.867~\text{GHz}\) and is strongly coupled to the external environment. All control pulses, readout tones, and sideband drives are delivered exclusively through the readout cavity via a strongly coupled port.
	
	\begin{table}[h] 
		\begin{tabular}{l|ccc}
			& \textbf{X (mm)} & \textbf{Y(mm)}& \textbf{Z(mm)} \\ \hline
			Readout cavity (6.8 GHz) & 26.6 & 5 & 40 \\
			Storage cavity (5 GHz)   & 45   & 5 & 40
		\end{tabular}
		\caption{Geometrical dimensions of the readout and storage cavities.}
		\label{tab:CavDims}
	\end{table}
	
	\subsection{The transmon qubit}
	
\begin{figure}[h]
		\centering
		\includegraphics[width=0.95\linewidth]{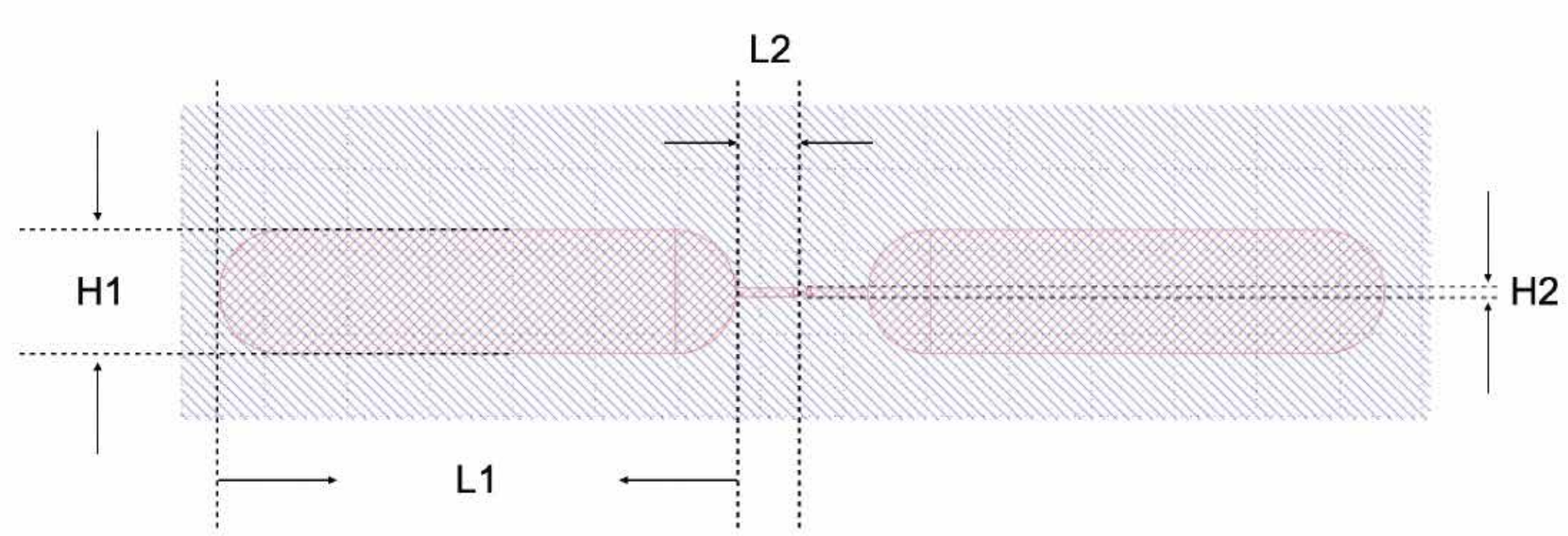}
		\caption{Design of the superconducting transmon qubit. The final dimensions used in the prototype were $L1 = 1$ mm, $H1 = 0.3$ mm, $L2 = 0.15$ mm, $H2 = 0.02$ mm. The Josephson junction is located at the center of the structure, connecting the two antenna pads.}
		\label{fig:Transmon_design}
	\end{figure}

	The transmon qubit was designed to couple dispersively to both the readout and storage cavities. Its geometry consists of two large antenna pads connected by a Josephson junction, as shown in Fig.~\ref{fig:Transmon_design}. To allow post-fabrication selection and optimization, eight transmon designs with varying pad dimensions were fabricated on a single wafer. Electromagnetic simulations predicted an anharmonicity of approximately 200~MHz and a storage–transmon Stark shift \(2\chi_\text{qs}\) on the order of 3~MHz. Experimentally, we measured a larger anharmonicity of $-272$~MHz and a Stark shift of \(2\chi_\text{qs}/2\pi = -4.6~\text{MHz}\), consistent with strong coupling to the storage cavity. 
	
	Fabrication proceeded in two steps. First, aluminum (Al) antenna pads were patterned using optical lithography and etched via reactive ion etching. Subsequently, the Josephson junction was fabricated using electron-beam lithography and the Dolan bridge technique, yielding an Al/AlO$_x$/Al trilayer junction. The resulting junctions exhibited critical currents corresponding to Josephson inductances of $L_J\sim 12$~nH and junction capacitances of $C_J\sim 3.5$~pF. After dicing, individual transmon chips were mounted into the cavity assembly. Further fabrication details are provided in \cite{dixit2021thesis}.

	\subsection{The wiring diagram}
	
\begin{figure}
		\centering
		\includegraphics[width=0.95\linewidth]{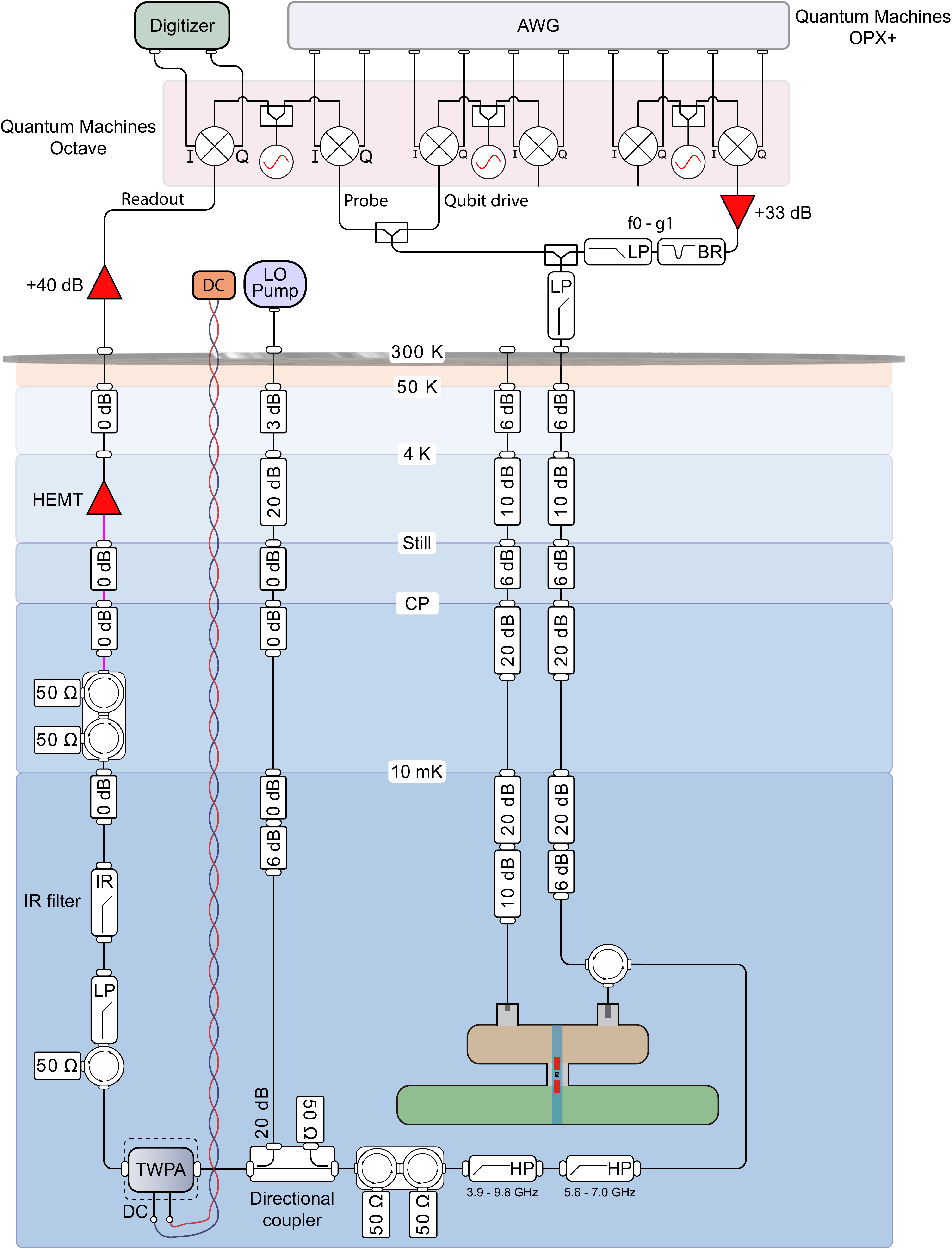}
		\caption{Schematic of the cryogenic wiring and control setup. The sample is mounted at the mixing chamber stage ($\approx 10$ mK). Input signals are attenuated at multiple temperature stages to suppress thermal noise. Output signals are amplified by a TWPA at the base temperature, a HEMT amplifier at the 4~K stage and Narda-Miteq amplifier at room temperature before being digitized by the OPX system.}
		\label{fig:wiring}
	\end{figure}

	The cavity assembly was thermally anchored to a copper plate mounted on the mixing chamber stage of a Bluefors dilution refrigerator. The device was enclosed in a shielded can to protect it from spurious magnetic noise. All measurements were performed at a base temperature of approximately 10~mK. A schematic of the cryogenic wiring and room-temperature control electronics is shown in Fig.~\ref{fig:wiring} (the copper plate is not shown).
	
	Microwave control and readout pulses were generated at room temperature using a Quantum Machines OPX+/Octave system, providing waveform synthesis over a 350~MHz bandwidth. The combined signals were attenuated at each stage of the dilution refrigerator before reaching the cavity, minimizing thermal noise.
	
	The readout signal was first amplified at the mixing chamber stage using a traveling-wave parametric amplifier (TWPA) from VTT Technical Research Centre of Finland, followed by further amplification at the 4~K stage using a high-electron-mobility transistor (HEMT) amplifier from Low Noise Factory. After exiting the refrigerator, the signal was amplified once more by a Narda-Miteq LNA-40-04000800-07-10P amplifier before being demodulated and digitized by the OPX+/Octave system.
	
	\section{Calibration of driving pulses}\label{calibration}
	
	\subsection{Basic qubit characterization}
	
	We experimentally characterized the transition frequencies of the transmon qubit up to the fourth excited state, from $\omega_{ge}$ to $\omega_{fh}$. The detailed parameters are summarized in Table \ref{tab:params}. Each transition serves a distinct experimental objective. 
	
	The fundamental transition $\omega_{ge}$ was used to calibrate $X_{\pi}$ and $X_{\pi/2}$ control pulses, which constitute the basic building blocks of the parity measurement protocol. The amplitude and waveform of these pulses are calibrated in Appendix~\ref{calibration}.4. At this frequency, we measured a qubit energy relaxation time of ($T_\text{1,q} = 16.7\,\mu\text{s}$) and a coherence time of ($T_\text{2,q} = 6.2\,\mu\text{s}$). 
	
	The second transition $\omega_{ef}$ was characterized to extract the qubit anharmonicity $\alpha = \omega_{ef} - \omega_{ge}$, a vital parameter for optimizing the waveform of the robust $X_{\pi/2}$ pulse. In addition, access to the $|f\rangle$ level is essential for photon injection into the storage cavity via the $\omega_{f0\text{-}g1}$ sideband transition; the calibration of this transition is described in Appendix~\ref{calibration}.2.
	
	Finally, the higher-order transition $\omega_{fh}$ was measured to verify sufficient spectral separation from all driven tones, ensuring that no spurious excitations occur during device operation. We also measure an effective qubit temperature of 72mK\cite{sultanov2021protocol}.
	
	\subsection{Photon injection}
	
	As discussed in Appendix~\ref{setup}, the storage cavity is not directly addressable by an external drive. Photon injection is therefore mediated by the transmon qubit, which couples dispersively to the storage mode. The injection protocol proceeds as follows. Starting from the ground state $|g, n\rangle$, the qubit is sequentially excited to the $|f\rangle$ state by applying resonant $\pi$ pulses on the $\omega_{ge}$ and $\omega_{ef}$ transitions. Subsequently, a driven sideband transition at frequency $\omega_{f0\text{-}g1}$ coherently transfers the excitation from the qubit to the storage cavity, implementing the state transfer
	$|f,n\rangle \rightarrow |g,n+1\rangle$, thereby increasing the photon number in the storage cavity by one.
	
	\begin{table}
		\caption{\label{tab:params} Summary of experimental device parameters.}
		\begin{ruledtabular}
			\begin{tabular}{lcc}
				\textrm{Device parameter} & \textrm{Symbol} & \textrm{Value} \\
				\colrule
				Qubit frequency $\vert g \rangle \rightarrow \vert e \rangle$ & $\omega_{ge}$ & $2 \pi ~\times$ 4.629 GHz \\
				Qubit frequency $\vert e \rangle \rightarrow \vert f \rangle$ & $\omega_{ef}$ & $2 \pi ~\times$ 4.357 GHz \\
				Qubit frequency $\vert f \rangle \rightarrow \vert h \rangle$ & $\omega_{fh}$ & $2 \pi ~\times$ 4.055 GHz \\
				Qubit anharmonicity & $\alpha$ & $-2 \pi ~\times$ 272 MHz \\
				Qubit decay time & $T_\text{\rm 1,q}$ & 16.7 $\mu$s \\
				Qubit decoherence time & $T_\text{\rm2,q}$ & 6.2 $\mu$s \\
				Qubit Temperature & $T_{\rm eff,q}$ & 72 mK \\
				\colrule
				Storage frequency & $\omega_\text{s}$ & $2 \pi ~\times$ 5.051 GHz \\
				Storage decay time & $T_\text{\rm 1,s}$ & 20.3 $\mu$s \\
				Storage-Qubit Stark shift  & $2\chi_\text{qs}$ & $-2 \pi ~\times$ 4.6 MHz \\
				Sideband transition & $\omega_{f0\text{-}g1}$ & $2 \pi ~\times$ 3.935 GHz \\
				\colrule
				Readout frequency & $\omega_\text{r}$ & $2 \pi ~\times$ 6.867 GHz \\
				Readout fidelity $\vert g \rangle$ \& $\vert e \rangle$ & $\mathcal{F}$ & 0.95739 \\
				\colrule
				Readout Pulse length & $T_\text{r}$ & 800 ns \\
				$X_{\pi / 2}$ gate pulse length & $T_\text{g}$ & 52 ns \\
				Delay time between $X_{\pi / 2}$ & $T_{\pi / 2}$ & 88 ns \\
				Waiting time between cycle & $T_\text{d}$ & 100 ns \\
				Parity sequence total length& $T_\text{t}$ & 1.092 $\mu$s \\
			\end{tabular}
		\end{ruledtabular}
	\end{table}
	
	\begin{figure}[t]
		\centering
		\includegraphics[width=\linewidth]{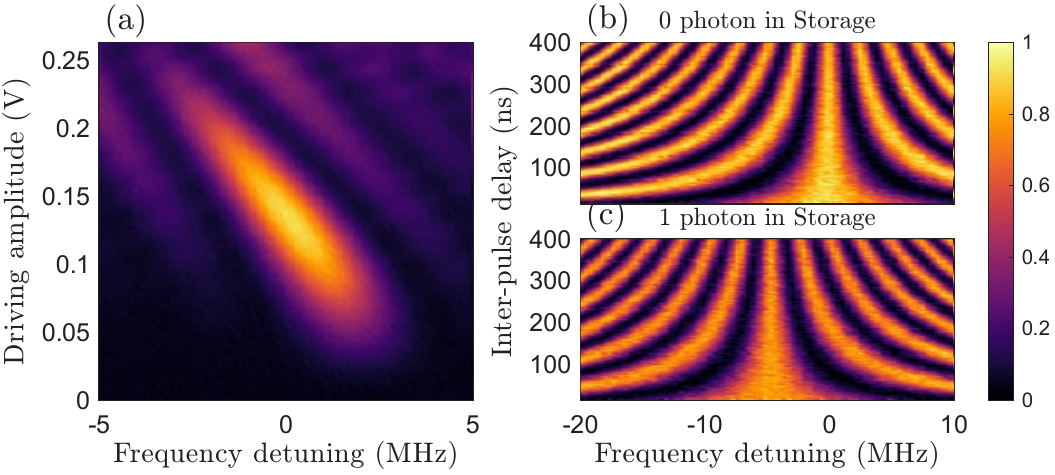}
		\caption{\textbf{Calibration of the $f0$–$g1$ sideband and the storage--transmon Stark shift.} \textbf{(a)} Two-dimensional map of the transmon $|f,0\rangle\!\to\!|g,1\rangle$ sideband response: the horizontal axis is the drive detuning from $\omega_{f0\text{-}g1}=2\pi\!\times\!3.935$~GHz and the vertical axis is the driving amplitude at the AWG output. The chevron pattern at small amplitude collapses into a single sideband resonance whose centre redshifts at large amplitude due to an AC Stark effect. \textbf{(b),(c)} Ramsey interferograms of the transmon $\omega_{\rm ge}$ transition with the storage cavity prepared in (b) the vacuum ($n=0$) and	(c) the single-photon Fock state ($n=1$). Horizontal axis: probe detuning from $\omega_{\rm ge}=2\pi\!\times\!4.629$~GHz; vertical axis: inter-pulse delay between the two $\pi/2$ pulses. Colour encodes the measured excited-state population $P_{\rm e}$ (scale at right). The horizontal shift between (b) and (c) yields $2\chi_{\rm qs}/2\pi=-4.6$~MHz.
		}
		\label{fig:f0g1}
	\end{figure}
	
	The sideband transition $\omega_{f0\text{-}g1}$ is not a first-order process and therefore requires relatively high drive power to achieve efficient population transfer. To this end, a +33~dB power amplifier (Mini-Circuits ZVE-3W-83+) was used to boost the drive amplitude. The measurement in Fig.~\ref{fig:f0g1}(a) spans a detuning range of $\pm 5$~MHz around $\omega_{f0\text{-}g1}=2\pi\!\times\!3.935$~GHz. At high drive amplitudes, a redshift of the effective transition frequency is observed, which we attribute to an AC Stark shift induced by the strong drive. Owing to the spectral proximity of $\omega_{f0\text{-}g1}$ to the qubit transition frequencies, a band-reject filter (Wainwright WTRCJV6-4000-5000-3-22-30SS) was installed after the amplifier to suppress spectral components near $\omega_{ge}$ and $\omega_{ef}$. This filtering prevents unintended excitation of higher qubit transitions during photon injection. The sideband frequency satisfies the relation $\omega_{f0\text{-}g1} = \omega_{ge} + \omega_{ef} - \omega_\text{s}$, from which we independently confirm the storage cavity frequency $\omega_\text{s} = 5.051$ GHz.
	
	Ramsey interferometry was utilized to infer the photon number in the storage cavity via the photon-number-dependent frequency shift of the transmon qubit. In the dispersive regime, the qubit transition frequency depends on the cavity photon occupation $n$ according to $\omega_{ge}(n) = \omega_{ge} + 2 \chi_\text{qs} n$, where $2\chi_\text{qs}$ is the storage–qubit Stark shift.
	Fig.~\ref{fig:f0g1}(b) and (c) show representative Ramsey interference fringes measured with the storage cavity prepared in the (b) $n=0$ and (c) $n=1$ photon states. The clear frequency offset between the two fringe patterns directly reflects the dispersive coupling between the qubit and the storage mode. Fitting the centre frequency of each Ramsey fringe pattern yields a separation of $2\chi_{\rm qs}/2\pi=-4.6$~MHz listed in Table~\ref{tab:params}.
	
	\subsection{Storage measurement}
	
\begin{figure}
		\centering
		\includegraphics[width=\linewidth]{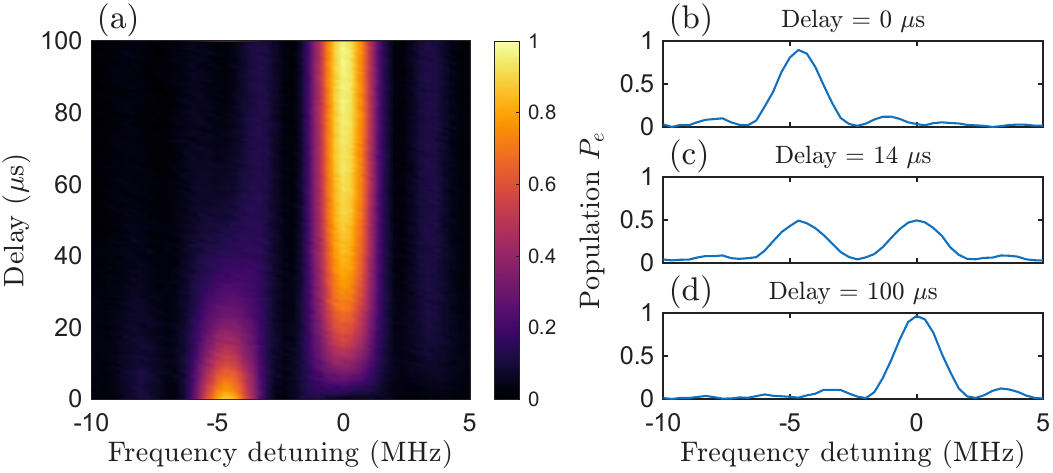}
		\caption{\textbf{Storage-cavity photon lifetime $T_{1,\rm s}$.} \textbf{(a)} Photon-number-resolved spectroscopy of the transmon $\omega_{ge}$ transition as a function of delay after photon injection. Horizontal axis: probe detuning from $\omega_{ge}=2\pi\!\times\!4.629$~GHz; vertical axis: delay; colour: excited-state population $P_{e}$ measured with a long	($720$~ns) $\pi$-probe (scale at right). The two bands at detunings $0$ and $2\chi_{\rm qs}/2\pi=-4.6$~MHz correspond to the storage Fock states $n=0$ and $n=1$, respectively. \textbf{(b)–(d)} Horizontal slices through (a) at delays $0$, $14$, and $100~\mu$s. The $n=1$ peak decays monotonically with delay while the $n=0$ peak grows correspondingly; an exponential fit to the $n=1$ population yields $T_{1,\rm s}=20.3~\mu$s.}
		\label{fig:T1s}
	\end{figure}

	Following photon injection, we characterized the single-photon lifetime of the storage cavity. As shown in Fig.~\ref{fig:T1s}, the photon population was monitored indirectly by tracking the photon-number-resolved qubit transition $\omega_{ge}(n=1)$ using spectroscopy. To spectrally resolve the $\omega_{ge}(n=0)$ and $\omega_{ge}(n=1)$ transitions, we employed a long-duration $\pi$ pulse (720 ns), thereby reducing the spectral bandwidth of the excitation to well below the $-4.6$~MHz Stark splitting calibrated in Appendix~\ref{calibration}.2. The population associated with the $n=1$ was observed to decay monotonically with increasing delay time, accompanied by a corresponding increase in the $n=0$ population. An exponential fit to the decay curve yields a storage photon lifetime of $T_\text{1,s} = 20.3~\mu\text{s}$. This lifetime sets the upper bound on the number of parity cycles that can be performed within the photon coherence window, $N\!\sim\!T_{1,\rm s}/T_{\rm t}$ (Sec.~\ref{sec:Intro}).
	
	\subsection{Drive pulse calibration}

	The parity measurement protocol, comprising two $X_{\pi/2}$ gates separated by a free-evolution interval $T_p$, forms the core of the photon detection scheme. Accurate calibration of both the gate amplitude and the timing parameters is therefore essential. The amplitude of the robust $X_{\pi/2}$ pulse was calibrated using a pseudo-identity sequence, $(X_{\pi/2})^{4n} = (-1)^n I,$ where the pulse amplitude was fine-tuned to maximize the return probability to the qubit ground state. This procedure is particularly sensitive to systematic over- or under-rotations and provides a reliable calibration of the pulse area.
	
\begin{figure}[t]
		\centering
		\includegraphics[width=0.9\linewidth]{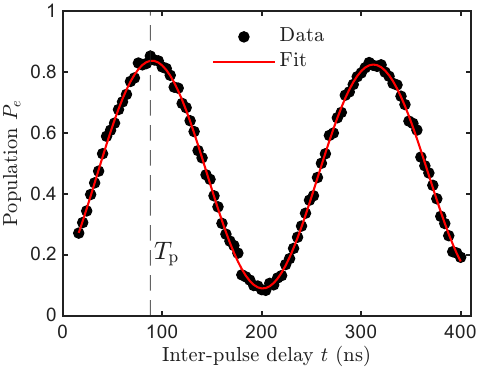}
		\caption{\textbf{Calibration of the parity-mapping delay $T_{\rm p}$.} A Ramsey-type scan is performed using the parity-measurement sequence [two $X_{\pi/2}$ pulses separated by $t$, see Fig.~\ref{fig:sequence}(c)] with the storage cavity prepared in the single-photon Fock state ($n=1$) using the protocol of Appendix~\ref{calibration}.2. The horizontal axis is the inter-pulse delay $t$ and the vertical axis is the excited-state population $P_{\rm e}$ measured after the second $X_{\pi/2}$ pulse. Black markers are data; the red curve is a fit of the form $P_{\rm e}(\tau)=A+B\cos(2\pi f t+\phi)$ with $f=4.5$~MHz, consistent with $|2\chi_{\rm qs}|/2\pi=4.6$~MHz from Appendix~\ref{calibration}.2. The first maximum, $t=T_{\rm p}=88$~ns, is selected as the parity-mapping interval.}
		\label{fig:rx90}
	\end{figure}

	Following amplitude calibration, we determine the free-evolution interval $T_{\rm p}$ used in stage 2 of the parity-measurement cycle [Sec.~\ref{sec:Methods}.A, Fig.~\ref{fig:sequence}(c)]. With the storage cavity prepared in $n=1$ via the protocol of Appendix~\ref{calibration}.2, we perform a Ramsey-type scan of the parity sequence as a function of delay (Fig.~\ref{fig:rx90}). The qubit population exhibits periodic oscillations, with the first maximum occurring at $T_\text{p} = 88$ ns. This value is slightly shorter than the theoretical estimate $T_\text{p} = 1/(2\chi_\text{qs}) \approx 109$ ns. The discrepancy arises from additional phase accumulation during the finite-duration ($T_{\pi/2} = 52$ ns) of the robust pulses, which is not captured in the instantaneous-pulse approximation.
	
	\subsection{Quantum Process Tomography}
	\label{sec:qpt}
	
	\begin{figure}
		\centering
		\includegraphics[width=\linewidth]{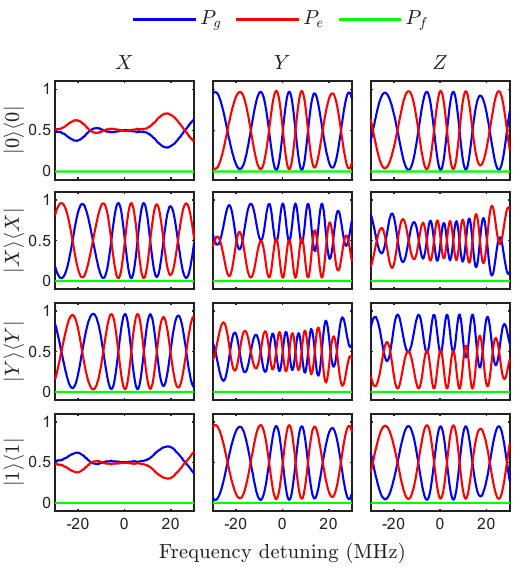}
		\caption{Quantum Process Tomography (QPT) of the robust $X_{\pi/2}$ gate under frequency detuning. The panels display the measured populations of the ground state ($P_g$, blue), excited state ($P_e$, red), and second excited state ($P_f$, green) as a function of frequency detuning from -30 to 30 MHz. The rows represent four distinct input states  $\{|0\rangle\langle 0|, |X\rangle\langle X|, |Y\rangle\langle Y|, |1\rangle\langle 1|\}$ while the columns correspond to the measurement bases Z, X, and Y. The characteristic oscillations observed in the populations arise from the phase accumulation induced by the frequency detuning during the robust $X_{\pi/2}$ gate operation.}
		\label{fig:qpt}
	\end{figure}
	
	Quantum Process Tomography (QPT) was employed as a diagnostic tool to characterize the experimentally implemented control gates, allowing for a quantitative evaluation of their fidelity with respect to the ideal target operation and a direct comparison between the standard Rectangular $X_{\pi/2}$ and Neural ODE–optimized Robust $X_{\pi/2}$ pulses. The QPT protocol begins by initializing the transmon qubit in the ground state $|g\rangle$. Four linearly independent input states are then prepared using the gate set $\{I, X_{\pi/2}, Y_{\pi/2}, X_{\pi}\}$, after which the gate under test (GUT) is applied. Finally, the state populations are measured in different basis: $\{|0\rangle\langle 0|, |X\rangle\langle X|, |Y\rangle\langle Y|, |1\rangle\langle 1|\}$.
	
	To ensure a fair comparison, both pulse types were characterized using the same gate duration of $T_{\pi/2} = 100$ ns, with measurements performed across a detuning range of $\Delta\!=\!(\omega_{\rm drive}-\omega_{\rm ge})/2\pi = \pm 30$ MHz. At finite detuning, phase accumulation during the gate results in rapid oscillations of the reconstructed process matrix, obscuring the intrinsic gate fidelity, as shown in Fig.~\ref{fig:qpt}. To remove this trivial effect, we apply a rotating-frame correction scheme.
	
	First, to separate coherent unitary errors from non-unitary decoherence processes, we fit the experimentally reconstructed process matrix $\chi_{\text{exp}}$ to the closest unitary process matrix, $\chi_{\text{fit}}$. Any single-qubit unitary operation $U$ can be expressed in terms of Pauli operators with coefficients $c_k$ as:
	\begin{equation}
		U = c_0 \sigma_0 - i \sum_{k \in \{x,y,z\}} c_k \sigma_k,
	\end{equation}
	where the diagonal elements of the corresponding process matrix are related to these coefficients by $\chi_{kk} = |c_k|^2$.
	
	Detuning-induced phase accumulation manifests as a rotation in the $X/Y$ subspace of these coefficients. The accumulated phase angle is given by:
	\begin{equation}
		\beta = 2\pi \delta (T_{\pi/2} + \Delta T)
	\end{equation}
	where $\Delta T$ accounts for the latency in the control electronics associated with oscillator frequency switching. To recover the intrinsic gate behavior, we apply an inverse rotation to the transverse coefficients:
	\begin{equation}
		c_x^{\text{corr}} + i c_y^{\text{corr}} = (c_x + i c_y) e^{-i\beta}
	\end{equation}
	This operation effectively transforms the data into a frame rotating at the detuned frequency, cancelling the trivial $z$-rotation induced by detuning. The gate performance is finally quantified by the Infidelity ($\mathcal{I}$), calculated using the corrected process matrix:
	\begin{equation}
		\mathcal{I} = 1 - \text{Tr}(\chi_{\text{fit}}^{\text{corr}} \cdot \chi_{\text{target}})
	\end{equation}
	where $\chi_{\text{target}}$ is the ideal process matrix for the $X_{\pi/2}$ gate. More details about the mathematical procedure can be found in \cite{kuzmanovic2025neural}.
	
	\subsection{Pulse-duration scaling of the Neural-ODE waveform}
	\label{sec:rescale}
	
	\begin{figure}[t]
		\centering
		\includegraphics[width=\linewidth]{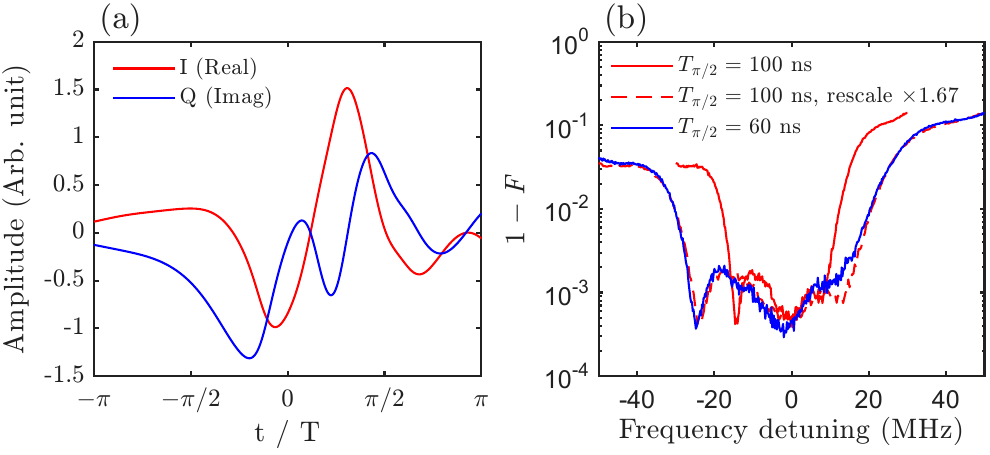}
		\caption{\textbf{Pulse-duration scaling of the Neural-ODE robust $X_{\pi/2}$ waveform.} \textbf{(a)} In-phase $I=\Omega_x$ (red) and quadrature $Q=\Omega_y$ (blue) components of the optimized envelope plotted as a function of the dimensionless time coordinate t/T used during Neural-ODE optimization; the physical pulse duration $T_{\pi/2}$ is obtained by mapping this coordinate onto a interval, so the same envelope can be replayed at any $T_{\pi/2}$ without re-optimization. \textbf{(b)} Gate infidelity $1-\mathcal{F}$ extracted from quantum process tomography (Appendix~\ref{sec:qpt}) as a function of the frequency detuning $\Delta\!=\!(\omega_{\rm drive}-\omega_{\rm ge})/2\pi$, for two instantiations of the envelope shown in (a): $T_{\pi/2}=100$~ns (red solid) and $T_{\pi/2}=60$~ns (blue solid).  Both pulses reach an on-resonance floor $1-\mathcal{F}\!\sim\!4\times 10^{-4}$. The red dashed curve shows the $T_{\pi/2}=100$~ns trace with its detuning axis rescaled by $100/60\!\approx\!1.67$; its near-perfect overlap with the blue trace is a direct experimental verification that the operational detuning window scales as $1/T_{\pi/2}$.}
		\label{fig:rescale}
	\end{figure}
	
	The Neural-ODE optimization is performed in a dimensionless time coordinate, so that the optimized in-phase and quadrature envelopes $\Omega_{x,y}$ [Fig.~\ref{fig:rescale}(a)] define a family of waveforms parameterised by a single physical duration $T_{\pi/2}$. A new physical pulse is obtained by replaying the same dimensionless envelope at a different sample points;
	no re-optimization is required\cite{kuzmanovic2025neural}.
	
	To verify this scaling experimentally we performed QPT on two instantiations of the same dimensionless envelope, $T_{\pi/2}=100$~ns and $T_{\pi/2}=60$~ns, processed with the rotating-frame correction of Sec.~\ref{sec:qpt}. The extracted infidelity is shown in Fig.~\ref{fig:rescale}(b). Both curves reach the same on-resonance floor $1-\mathcal{F}\!\sim\!4\times 10^{-4}$, confirming that the rescaling preserves the on-resonance gate fidelity.	Rescaling the detuning axis of the $T_{\pi/2}=100$~ns trace by $T_{\pi/2}^{(100)}/T_{\pi/2}^{(60)}=100/60\!\approx\!1.67$ [red dashed curve in Fig.~\ref{fig:rescale}(b)] brings it into near-perfect agreement with the directly measured $T_{\pi/2}=60$~ns trace across the entire $\pm 30$~MHz measurement window. This collapse provides a verification that the operational detuning window in physical units obeys the predicted $1/T_{\pi/2}$ scaling, and that the dimensionless waveform inherited from the original Neural-ODE optimization remains robust when rescaled to the duration actually used in the parity-measurement protocol.
	
	The lower bound on $T_{\pi/2}$ in our setup is set by the time resolution of Quantum Machines OPX+: with a sample rate of $1$~GS/s, the $T_{\pi/2}=52$~ns instance is rendered with $52$ samples across the pulse, which is sufficient to resolve the waveform of the optimized envelope. Below $52$ samples, the rescaled waveform is no longer faithfully reproduced; this hardware limit, not the optimization itself, fixes the shortest $T_{\pi/2}$ accessible in the present work.
	
	\section{Data Analysis and Statistical Framework}
	\label{Framework}
	
	\subsection{Notation}
	\label{sec:Cnotation}
	Two distinct counts both denoted by a capital $N$ appear in the	analysis of this Appendix. We adopt the convention used throughout the main text: $N=21$ is the number of consecutive parity measurements composing one sequence (a single execution of the protocol of Fig.~\ref{fig:sequence}(c) after one initialization), fixed by the cycle time $T_{\rm t}=1.092~\mu$s. $N_{\rm meas}=N_{\rm inj}=N_{\rm bkg}=10{,}000$ is the number of independent sequences in each of the injection and no-injection runs of Sec.~\ref{sec:Benchmark}.A. Likelihood ratios $\lambda_i$	[Eq.~\eqref{eq:lambda}] are indexed by the sequence label $i$ within either run; the threshold $\lambda_{\rm th}$ acts at the run level on the collection $\{\lambda_i\}$.
	
	\subsection{Statistical Evaluation of Detector Performance}
	\begin{figure}[t]
		\centering
		\includegraphics[width=\linewidth]{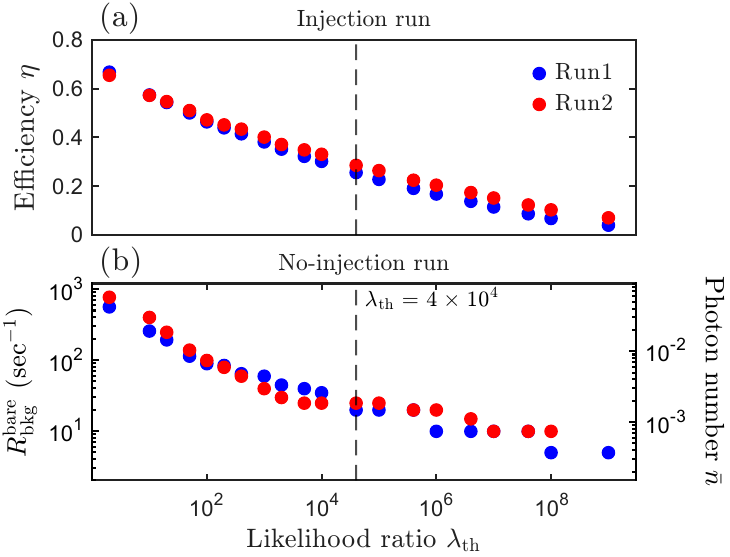}
		\caption{\textbf{Detector performance characterization}. \textbf{(a)} Detection efficiency $\eta$ as a function of the likelihood ratio $\lambda_{\rm th}$ and \textbf{(b)}Bare background rate $R_\text{bkg}^\text{bare}$ (left axis) and the corresponding average background photon number $\bar{n}$ (right axis) versus $\lambda_{\rm th}$. The markers (blue/red) represent independent experimental runs. The dashed vertical line indicates the chosen operational threshold $\lambda_\text{th} = 4 \times 10^{4}$ used to define the background-dominated regime, which maintains a detection efficiency $\eta \approx 0.27$ while suppressing the bare background rate $R_\text{bkg}^\text{bare} \approx 22~\text{sec}^{-1}$ and the average background photon number $\bar{n_c} \approx 1.7 \times 10^{-3}$.}
		\label{fig:BackgroundRates}
	\end{figure}
	
	To quantify the detector response we adopt two operational figures of merit, defined entirely from the sequence-level likelihood ratios $\{\lambda_i\}$ of Eq.~\eqref{eq:lambda}. For a threshold $\lambda_{\rm th}$, the number of triggered sequences in the injection and no-injection runs are
	\begin{align}
		N_{\rm det}(\lambda_{\rm th}) &= \sum_{i=1}^{N_{\rm inj}} \Theta\bigl(\lambda_i-\lambda_{\rm th}\bigr),\label{eq:ndet}\\
		N_{\rm false}(\lambda_{\rm th}) &= \sum_{j=1}^{N_{\rm bkg}} \Theta\bigl(\lambda_j-\lambda_{\rm th}\bigr), \label{eq:nfalse}
	\end{align}
	where $\Theta$ is the Heaviside step function. The detection efficiency	under controlled injection is
	\begin{align}
		\eta(\lambda_{\rm th}) &= \frac{N_{\rm det}(\lambda_{\rm th})}{N_{\rm inj}}, \label{eq:eta}
	\end{align}
	i.e.\ the conditional probability that a sequence containing the deterministically prepared single photon yields $\lambda\!\geq\!\lambda_{\rm th}$.
	
    The corresponding figure of merit in the no-injection run is the bare per-sequence trigger probability $N_{\rm false}(\lambda_{\rm th})/N_{\rm bkg}$. Since each measurement sequence duration is limited by the storage lifetime $T_{1,s}$, we define the bare background trigger rate as
    \begin{align}
		R_{\rm bkg}^{\rm bare}(\lambda_{\rm th}) &= \frac{N_{\rm false}(\lambda_{\rm th})}{N_{\rm bkg} T_{1,\rm s}}, \label{eq:Rbkg}
	\end{align}
    The bare trigger probability is not, however, equal to the residual photon occupation of the storage cavity, because it is filtered by the same detector response $\eta(\lambda_{\rm th})$ that also governs the injection run. Removing this common response yields the calibrated residual occupation
	\begin{align}
		\bar n_c (\lambda_{\rm th}) &= \frac{N_{\rm false}(\lambda_{\rm th})}{N_{\rm det}(\lambda_{\rm th})} = \frac{1}{\eta(\lambda_{\rm th})} \frac{N_{\rm false}(\lambda_{\rm th})}{N_{\rm bkg}}, \label{eq:nbar}
	\end{align}
	where the last equality uses $N_{\rm inj}=N_{\rm bkg}=10{,}000$. We interpret $\bar n_c$ as the time-averaged equivalent photon occupation of the storage cavity over one parity-measurement cycle: under the same response function applied to the injected signal, any residual trigger in the no-injection run is converted to a photon count. The corresponding  efficiency-corrected background photon rate is
	\begin{align}
		R_{\rm bkg}^{\rm corr}(\lambda_{\rm th}) &= \frac{\bar n_c (\lambda_{\rm th})}{T_{1,\rm s}}, \label{eq:Rbkg}
	\end{align}
    the steady-state rate at which incoherent background photons populate and leak out of the stor
	
	The plateau structure of $\bar n_c (\lambda_{\rm th})$ at intermediate	$\lambda_{\rm th}$ acquires a transparent interpretation under this	definition. Let $p_1(\lambda_{\rm th})$ denote the conditional	probability that a sequence with a photon present at $t=0$ yields $\lambda\!\geq\!\lambda_{\rm th}$. Once $\lambda_{\rm th}$ is large	enough to suppress signal-tail mis-classifications in both runs,
	\begin{equation*}
		\frac{N_{\rm det}}{N_{\rm inj}} \to p_1(\lambda_{\rm th}),
		\qquad
		\frac{N_{\rm false}}{N_{\rm bkg}} \to \bar n_{\rm phys}\,p_1(\lambda_{\rm th}),
	\end{equation*}
	so the ratio in Eq.~\eqref{eq:nbar} converges to the intrinsic occupation $\bar n_{\rm phys}$ with the response factor explicitly	cancelled. The operational threshold $\lambda_{\rm th}=4\times 10^{4}$	(Sec.~\ref{sec:Benchmark}.A, Fig.~\ref{fig:BackgroundRates}) is placed
	at the onset of this plateau, where further increases of $\lambda_{\rm th}$ no longer change $\bar n_c$ within statistical uncertainty.
	
	At the operational threshold $\lambda_{\rm th}=4\times 10^{4}$, the	no-injection runs yield $R_\text{bkg}^{\rm bare} \approx 22~\text{sec}^{-1}$ which correspond to the per-run averaged residual cavity occupation
	\begin{equation}
		\bar n_{c} \;\equiv\; \frac{1}{\eta(\lambda_{\rm th})}\frac{N_{\rm false}(\lambda_{\rm th})}{N_{\rm bkg}}\;\approx\; 1.7\times 10^{-3},
		\label{eq:nbar_c}
	\end{equation}
	and equivalently $R_{\rm bkg}^{\rm corr}=\bar n_{c}/T_{1,\rm s}\!\approx\!80~\text{sec}^{-1}$ with $T_{1,\rm s}=20.3~\mu$s. The physical decomposition of $\bar n_{c}$ into a transmon-mediated dispersive-dressing contribution and the residual extraneous-bath contribution is treated in Appendix~\ref{sec:nbar_sources}.
	
	\subsection{Source of residual photons}
	\label{sec:nbar_sources}
	The residual cavity occupation reported in Eq.~\eqref{eq:nbar_c} corresponds, under the Bose--Einstein relation	$\bar n_{c}=\bigl[\exp(\hbar\omega_{\rm s}/k_{\rm B}T_{\rm eff,s})-1\bigr]^{-1}$	with $\omega_{\rm s}/2\pi=5.051$~GHz ($\hbar\omega_{\rm s}/k_{\rm B}\!\approx\!242$~mK), to an effective storage-mode temperature
	\begin{equation}
		T_{\rm eff,s}\;=\;\frac{\hbar\omega_{\rm s}/k_{\rm B}}{\ln\bigl(1+1/\bar n_{c}\bigr)}\;\approx\;38~\mathrm{mK}.
		\label{eq:Tceff}
	\end{equation}
	This value sits between the mixing-chamber stage of the dilution refrigerator ($T_{\rm MC}\!\approx\!12$~mK, measured at the cold plate) and the transmon effective temperature $T_{\rm eff,q}\!\approx\!72$~mK inferred from the residual qubit population in Table~\ref{tab:params}. The hierarchy $T_{\rm MC}<T_{\rm eff,s}<T_{\rm eff,q}$ is not a thermodynamic	contradiction: in the dispersive limit the cavity--qubit interaction $H_{\rm int}/\hbar = 2\chi_{\rm qs}\,a^{\dagger}a\,\sigma_{z}/2$ contains no Jaynes--Cummings exchange term $g_{\rm qs}(\sigma_{+}a + \sigma_{-}a^{\dagger})$, so the two modes	do not thermalize to each other and equilibrate independently with their respective dissipation channels\cite{dixit2021searching}. The storage cavity, a passive 3D aluminum resonator without a Josephson element, is in particular not directly heated by the quasiparticle bath that dominates the transmon thermal population, allowing $T_{\rm eff,s}\!<\!T_{\rm eff,q}$. The remaining offset $T_{\rm eff,s}-T_{\rm MC}\!\approx\!26$~mK is attributed to extraneous thermal channels, the dominant of which we estimate in what follows.
	
	The transmon population leaks into a residual cavity excitation	through the small dispersive dressing between the bare states $|e,0\rangle$ and $|g,1\rangle$. Within the single-excitation manifold of the Jaynes--Cummings Hamiltonian, an exact two-by-two diagonalization gives the dressed eigenstates
	\begin{align}
		|\widetilde{e}\rangle &= \cos\theta\,|e,0\rangle + \sin\theta\,|g,1\rangle,\\
		|\widetilde{1}\rangle &= -\sin\theta\,|e,0\rangle + \cos\theta\,|g,1\rangle,
	\end{align}
	with mixing angle determined by $\tan 2\theta = 2g_{\rm qs}/\Delta$, where $\Delta\!\equiv\!\omega_{\rm q}-\omega_{\rm s}$ is the qubit--storage detuning. In the dispersive limit $g_{\rm qs}\!\ll\!|\Delta|$ this reduces to
	\begin{equation}
		\tan\theta \;\simeq\; \frac{g_{\rm qs}}{|\Delta|}.
		\label{eq:tan_theta}
	\end{equation}
	A quasiparticle-induced tunneling event prepares the transmon in the bare state $|e,0\rangle$; projected onto the dressed basis, this event is registered by the parity-measurement protocol as a cavity excitation with probability $|\langle\widetilde{1}|e,0\rangle|^{2}=\sin^{2}\theta\simeq (g_{\rm qs}/\Delta)^{2}$. The transmon-mediated contribution to the	cavity occupation is therefore
	\begin{equation}
		\bar n_{c}^{q}\;=\; \bar n_{q}\,(g_{\rm qs}/\Delta)^{2},
		\label{eq:ncq}
	\end{equation}
	with $\bar n_{q}$ the transmon residual occupation.
	
	For our parameters $\omega_{\rm q}/2\pi=4.629$~GHz, $\omega_{\rm s}/2\pi=5.051$~GHz, $\alpha_{\rm q}/2\pi=-272$~MHz,	and $\chi_{\rm qs}/2\pi=-2.3$~MHz (Table~\ref{tab:params}), the	transmon--storage coupling inferred from the dispersive shift $\chi_{\rm qs}=g_{\rm qs}^{2}\alpha_{\rm q}/[\Delta(\Delta+\alpha_{\rm q})]$ is $g_{\rm qs}/2\pi\!\simeq\!49.8$~MHz, giving	$(g_{\rm qs}/\Delta)^{2}\!\simeq\!1.4\times 10^{-2}$\cite{PhysRevA.76.042319}. With $T_{\rm eff,q}\!\approx\!72$~mK at $\omega_{\rm q}/2\pi=4.629$~GHz ($\hbar\omega_{\rm q}/k_{\rm B}\!\approx\!222$~mK), the residual	transmon occupation is $\bar n_{q}\!\approx\!4.8\times 10^{-2}$, and Eq.~\eqref{eq:ncq} yields
	\begin{equation}
		\bar n_{c}^{q}\;\simeq\;4.8\!\times\!10^{-2}\times 1.4\!\times\!10^{-2}\;\simeq\;6.7\times 10^{-4},
		\label{eq:ncq_value}
	\end{equation}
	accounting for $\sim\!39\%$ of the measured residual occupation $\bar n_{c}\!\approx\!1.7\times 10^{-3}$ [Eq.~\eqref{eq:nbar_c}]. The remaining $\sim\!61\%$ must be attributed to extraneous thermal channels not mediated by the qubit, including blackbody radiation from higher temperature stages of the dilution refrigerator, insufficient attenuation or thermalization of the input lines, and back-action noise from the readout	amplification chain\cite{dixit2021searching}.
	
	The detector therefore sits in an intermediate regime in which the two contributions are of comparable magnitude. Three limiting scenarios are worth distinguishing: $\bar n_{c}^{q}/\bar n_{c}\!\gtrsim\!50\%$ would indicate that qubit-side improvements (active reset, quasiparticle trapping, gap engineering) provide the dominant lever for reducing $\bar n_{c}$; $\bar n_{c}^{q}/\bar n_{c}\!\ll\!10\%$ would indicate that the environment-side improvements alone (infrared filtering, cryogenic attenuation, magnetic and electromagnetic shielding) would suffice; the present value $\sim\!39\%$ places the two routes on equal footing, with the engineering response discussed in Sec.~\ref{sec:Discussion}.
	
	\subsection{Dark Photon Model}
	
	The expected number of detected events induced by massive dark photons is given by:
	\begin{equation}
		N_\text{DM} = \frac{\epsilon^{2}\rho_\text{DM}Q_\text{DM}Q_\text{s}GV}{\omega_\text{s}} N_\text{meas}
	\end{equation}
	where $\epsilon$ is the kinetic mixing angle and $\rho_{DM}=0.45~\mathrm{GeV/cm^{3}}$ is the local dark matter density assumed from the Standard Halo Model. The factor $Q_\text{DM}\approx10^{6}$ corresponds to the expected dark matter signal linewidth, while $Q_\text{s}$ and $\omega_\text{s}$ are the quality factor and angular frequency of the storage cavity mode, respectively, with $T_\text{1,s} = Q_\text{s}/\omega_\text{s}$. The geometric form factor $G$ accounts for the overlap between the cavity electromagnetic mode and the effective electric field induced by the dark photon. Assuming the TE$_{101}$ mode is polarized along the $z$-axis, $G$ reduces to a one-dimensional overlap of the mode profile with a uniform external field:
	\begin{equation}
		G \;=\; \frac{1}{3}\,\frac{\bigl|\!\int_{V}\! E_{z}\,dV\bigr|^{2}}{V\!\int_{V}\!|E|^{2}\,dV}\;=\; \frac{1}{3}\,\frac{64}{\pi^{4}}\,,
		\label{eq:Gfactor}
	\end{equation}
	where the factor $1/3$ arises from the isotropic time-average over the random polarization direction of the dark photon, and the analytic value $64/\pi^{4}$ follows from the TE$_{101}$ mode profile	$E_{z}\propto\sin(\pi x/a)\sin(\pi z/c)$ of a rectangular cavity of	dimensions $a\!\times\!b\!\times\!c$ in Table \ref{tab:CavDims}. The effective cavity volume is $V = 9~\mathrm{cm}^3$.
	
	\subsection{Confidence Level Calculation}
	\label{sec:CL}
	
	\begin{table}[b]
		\caption{\label{tab:summary} Exclusion limits on the dark photon mixing angle at 95\% C.L. for $\rho_\text{DM}=0.45~\mathrm{GeV/cm^3}$, obtained from Runs 1, 2 and their combined analysis using the quantum sensor operated as a single-photon counter.}
		\begin{ruledtabular}
			\begin{tabular}{lccc}
				\textbf{Parameter} & \textbf{Run 1} & \textbf{Run 2} & \textbf{Runs 1 + 2} \\
				\colrule
				$f_\text{s}$ (GHz) & 5.051 & 5.051 & 5.051 \\
				$Q_\text{s}$ & 634727 (0.6M) & 634727 (0.6M) & 634727 (0.6M) \\
				$N_\text{DM}$ & 4 & 5 & 4 - 5 \\
				$N_\text{meas}$ & 10000 & 10000 & 20000 \\
				Efficiency $\eta$ & 0.257 & 0.286 & 0.257 - 0.286 \\
				$n^\text{CL}$ & 0.00358 & 0.00368 & 0.00261 \\
				$\epsilon^\text{CL}$ & $ 1.2217 \cdot 10^{-14}$ & $ 1.2387 \cdot 10^{-14}$ & $ 1.0446 \cdot 10^{-14}$ \\
			\end{tabular}
		\end{ruledtabular}
	\end{table}
	
	The statistical inference on the kinetic mixing angle $\epsilon$ uses only the trigger counts above the operational threshold $\lambda_{\mathrm{th}}$ (Sec.~\ref{Framework}), without further spectral or temporal cuts.
	
	We model the outcome of a single measurement run as a binomial process with $N_{\mathrm{meas}}$ independent trials and per-trial success probability $p(\epsilon)$. The probability mass function is
	\begin{equation}
		B\bigl(k\,\big|\,N_{\mathrm{meas}},\,p\bigr)\;=\;\binom{N_{\mathrm{meas}}}{k}\,p^{k}\bigl(1-p\bigr)^{N_{\mathrm{meas}}-k},
		\label{eq:binom-pmf}
	\end{equation}
	where, for the dark-photon hypothesis,
	\begin{equation}
		p(\epsilon)\;=\;\frac{\eta(\lambda_{\mathrm{th}})\,N_{\mathrm{DM}}(\epsilon)}{N_{\mathrm{meas}}}
		\;=\;\eta(\lambda_{\mathrm{th}})\,\frac{\epsilon^{2}\rho_{\mathrm{DM}}Q_{\mathrm{DM}}Q_{s}GV}{\omega_{s}},
		\label{eq:p-eps}
	\end{equation}
	with $\eta(\lambda_{\mathrm{th}})$ the detection efficiency defined in Eq.~\eqref{eq:eta} at the operating threshold $\lambda_{\mathrm{th}}$. The cumulative probability of observing no more than
	$N_{\mathrm{obs}}$ trigger events out of $N_{\mathrm{meas}}$ trials is then
	\begin{equation}
		P\bigl(\le\!N_{\mathrm{obs}}\,\big|\,\epsilon\bigr)\;=\;\sum_{k=0}^{N_{\mathrm{obs}}}\!\binom{N_{\mathrm{meas}}}{k}\,p(\epsilon)^{k}\bigl(1-p(\epsilon)\bigr)^{N_{\mathrm{meas}}-k},
		\label{eq:cdf}
	\end{equation}
	where $N_{\mathrm{obs}}\equiv N_{\mathrm{false}}(\lambda_{\mathrm{th}})$ [Eq.~\eqref{eq:nfalse}] is the trigger count of the no-injection run. Because no background subtraction is applied, the resulting bound on $\epsilon$ is conservative.
	
	To combine the two independent data takings (Run 1, Run 2), we treat them as statistically independent experiments with respective inputs $\{N_{\mathrm{meas},i},\eta_{i},N_{\mathrm{obs},i}\}_{i=1,2}$. Denoting the per-run CDFs by $P_{i}(\epsilon)$, the joint upper-tail probability factorises as
	\begin{equation}
		P_{\mathrm{combined}}(\epsilon)\;=\; P_{1}(\epsilon)\,P_{2}(\epsilon),
		\label{eq:Pcomb}
	\end{equation}
	and the 95\% C.L. exclusion limit is the smallest $\epsilon$ for which
	\begin{equation}
		P_{\mathrm{combined}}(\epsilon^{95\%}) \;=\; 0.05.
		\label{eq:eps95}
	\end{equation}
	Equivalently, in terms of the signal-induced mean occupation per cycle $\bar n_{\mathrm{sig}}(\epsilon)\equiv N_{\mathrm{DM}}(\epsilon)/N_{\mathrm{meas}}$, the bound can be reported as $\bar n_{\mathrm{sig}}^{95\%}$ (Table~\ref{tab:summary}). The numerical results in Table~\ref{tab:summary} are obtained by numerically inverting Eq.~\eqref{eq:eps95}; the combined dataset yields $\epsilon^{95\%}=1.04\times 10^{-14}$ at $\omega_{s}/2\pi = 5.051$ GHz.
	
\end{document}